\documentclass[useAMS,usenatbib]{mn2e}
\usepackage{graphicx} 
\usepackage{amsmath} 
\usepackage{color}
\usepackage{float}
\usepackage{caption}
\usepackage{placeins}
\usepackage{lipsum}
\usepackage{multicol}
\def\be{\begin{equation}} 
\def\ee{\end{equation}}

\def\kms{\,{\rm {km\, s^{-1}}}} 
\def\msun{{\Msun}}

\def\gsim{\lower.5ex\hbox{\gtsima}} 
\def\lsim{\lower.5ex\hbox{\ltsima}} \def\gtsima{$\; \buildrel > \over 
\sim \;$} \def\ltsima{$\; \buildrel < \over \sim \;$} \def\prosima{$\; 
\buildrel \propto \over \sim \;$} \def\gsim{\lower.5ex\hbox{\gtsima}} 
\def\lsim{\lower.5ex\hbox{\ltsima}} 
\def\simgt{\lower.5ex\hbox{\gtsima}} 
\def\simlt{\lower.5ex\hbox{\ltsima}} 
\def\simpr{\lower.5ex\hbox{\prosima}} \def\la{\lsim} \def\ga{\gsim} 
  
 \def\gtsima{$\; \buildrel > \over \sim \;$} 
\def\ltsima{$\; \buildrel < \over \sim \;$} 
\def\gsim{\lower.5ex\hbox{\gtsima}} 
\def\lsim{\lower.5ex\hbox{\ltsima}} 
\def\simgt{\lower.5ex\hbox{\gtsima}} 
\def\simlt{\lower.5ex\hbox{\ltsima}} 
\def\simpr{\lower.5ex\hbox{\prosima}} 
\def\la{\lsim} 
\def\ga{\gsim}

\def\msun{\,{\rm \Msun}}

\def\E3{{\cal E}_{\rm g}^{III}}

\def\Msun{\rm M_\odot}
\def\mbh{\rm M_{bh}}
\def\Zsun{\rm Z_\odot}

\def\Msun{\rm M_\odot}
\def\myr{\rm Myr}

\def\Zsun{\rm Z_\odot}
\def\M*{M_*}
\def\mbh{M_{bh}}
\def\Z*{Z_*}
\def\L*{L_*}
\def\muv{\rm M_{UV}}

\def\fws{f_*^w}
\def\luvs{L_*^{UV}}
\def\luvtot{L_{tot}^{UV}}
\def\fwb{f_{bh}^w }
\def\luvb{L_{bh}^{UV}}

\def\fs{f_*}
\def\fej{f_*^{ej}}
\def\feff{f_*^{eff}}

\def\der{{\rm d}}

\def \mges{M_*^{ge}}
\def \mgfs{M_*^{gf}}

\def \mgfb{M_{bh}^{gf}}
\def\faccb{f_{bh}^{ac}}
\def\maccb{M_{bh}^{ac}}
\def\med{M_{ed}}
\def\fed{f_{ed}}
\def\mcritb{M_{bh}^{crit}}
\def\mdmsa{M_{dm}^{sa}}
\def\mgsa{M_{g}^{sa}}
\def\myrs{\rm Myrs}

\title[Galaxies, BHs \& GWs]{The hierarchical assembly of galaxies and black holes in the first billion years: predictions for the era of gravitational wave astronomy} 
\author[Dayal et al.]{Pratika Dayal$^{1}$\thanks{p.dayal@rug.nl}, Elena M. Rossi$^2$, Banafsheh Shiralilou$^2$, Olmo Piana$^1$, 
\newauthor
Tirthankar Roy Choudhury$^3$ \& Marta Volonteri$^4$\\ 
$^{{1}}$ Kapteyn Astronomical Institute, University of Groningen, P.O. Box 800, 9700 AV Groningen, The Netherlands \\
$^{2}$  Leiden University, Oort Building, Niels Bohrweg 2, NL-2333 CA Leiden, The Netherlands\\
$^3$ National Centre for Radio Astrophysics, Tata Institute of Fundamental Research, Pune 411007, India\\
$^4$ Sorbonne Universites, UPMC Univ Paris 6 et CNRS, UMR 7095, Institut d'Astrophysique de Paris, 98 bis bd Arago, 75014 Paris, France \\
}

\begin{document} 
 
\date{} 

\maketitle

\begin{abstract}
In this work we include black hole (BH) seeding, growth and feedback into our semi-analytic galaxy formation model, {\it Delphi}. Our model now fully tracks the, accretion- and merger-driven, hierarchical assembly of the dark matter halo, gas, stellar and BH masses of high-redshift ($z \gsim 5$) galaxies. We explore a number of physical scenarios that include: {\it (i)} two types of BH seeds (stellar and those from Direct Collapse BH; DCBH); {\it (ii)} the impact of reionization; and {\it (iii)} the impact of instantaneous versus delayed galaxy mergers on the baryonic growth. Using a minimal set of mass- and $z$-independent free parameters associated with star formation and BH growth, and their associated feedback, and including suppressed BH growth in lower-mass galaxies, we show that our model successfully reproduces all available data sets for early galaxies and quasars. While both reionization and delayed galaxy mergers have no sensible impact on the evolving ultra-violet luminosity function, the impact of the former dominates in determining the stellar mass density for observed galaxies as well as the BH mass function. We then use this model to predict the {\it LISA} detectability of merger events at high-$z$. As expected, mergers of stellar BHs dominate the merger rates for all scenarios and our model predicts an expected upper limit of about 20 mergers using instantaneous merging and no reionization feedback over the 4-year mission duration. Including the impact of delayed mergers and reionization feedback provides about 12 events over the same observational time-scale.

\end{abstract}


\section{Introduction}
The detection of Gravitational Waves (GWs), from merging stellar Black Hole (BH) binaries and recently from the mergers of binary neutron stars, have opened a new observable window onto the low-$z$ Universe. Over the next decade, we expect the Laser Interferometer Space Antenna ({\it LISA}) to detect GW signals from merging massive ($10^{4}-10^{6} \msun$) black hole binaries at the centre of galaxies from redshifts as high as $z \sim 20$, if such BHs are already in place at those early epochs. These observations perfectly complement galaxy surveys, using the {\it Hubble} and Subaru telescopes and the forthcoming James Webb Space telescope ({\it JWST}) and the European Extremely Large telescope ({\it E-ELT}), that aim at yielding tantalising glimpses of the formation of the earliest galaxies in the era of cosmic dawn \citep[for a recent review see][]{dayal2018}. 

Predictions for such observations naturally require theoretical models that can consistently and simultaneously reproduce existing galaxy and BH observations before predictions are made for higher redshifts. In this work, we introduce BH seeding, growth and feedback into the {\it Delphi} semi-analytic model of galaxy formation. This model now fully tracks the hierarchical assembly of the dark matter, baryonic and BH components of early galaxies including the impact of feedback associated with star-formation, BHs and reionization. We use the results of our model to forecast the BH merger rate and the associated GW signature for {\it LISA} between $z \sim 5-20$. This redshift window covers the formation epoch of supermassive black hole (SMBH) seeds, whose nature and properties are currently outstanding questions. Three main formation channels are currently discussed (see also Section \ref{bhseeds}) that involve: {\it (i)} the first generation of massive metal-free Population III stars \citep[PopIII; e.g.][]{2001ApJ...551L..27M}; {\it (ii)} the monolithic collapse of gas in assembling protogalaxy \citep[e.g.,][]{loeb1994}; and {\it (iii)} the core collapse of the first ultra-dense nuclear star clusters \citep[e.g.,][]{2009ApJ...694..302D}. These channels differ in terms of the expected seed mass and ``birth" rate with PopIII remnants being the lightest ($\sim 100 \msun$) and most frequent and BHs resulting from gas collapse (Direct Collapse Black Holes; DCBHs) being the most massive and rarest \citep[see][for recent reviews]{2016PASA...33...51L,2018arXiv180706155H}. Therefore, catching the GW signals from merging BHs between $z \sim 5-20$ can shed unique insights on SMBH infancy \citep{2018arXiv180706967C}. Electromagnetic searches of SMBH seeds probe the fraction of the SMBH population in an active phase and provide their luminosity, therefore shedding light on the combination of their masses and accretion rates. An alternative but still indirect measure of the SMBH mass, which is one of quantities needed to test seed formation scenarios, can be estimated with spectroscopic observations. On the other hand, {\it LISA} can detect and directly provide the masses and spins for quiescent SMBHs, as long as they are in a coalescing binary. Therefore these two approaches are truly complementary and both are necessary in order to obtain a full view of the SMBH seed population in the early Universe.

In this paper we focus on BH-BH mergers that ensue after the merger of two galaxies, both hosting a central black hole, rather than the merger of SMBHs that are born in a binary \citep{2018MNRAS.479L..23H}. While such calculations have been carried out by previous works \citep[e.g.][]{sesana2007, 2011PhRvD..83d4036S,barausse2012}, the strength of our model lies, both, in the minimal set of free parameters used for star formation and BHs (and their associated feedback) as well as the number of physical scenarios explored that include: {\it (i)} two types of BH seeds (stellar and those from Direct Collapse BH; DCBH); {\it (ii)} the impact of reionization; and {\it (iii)} the impact of instantaneous versus delayed galaxy mergers on the baryonic growth of galaxies. Our model matches the key observables both for high-$z$ Lyman break galaxies (LBGs)- including the Ultra-violet luminosity function (UVLF), the stellar mass function (SMF), the stellar mass density (SMD), stellar mass-halo mass relations and the mass-to-light (M/L) ratios- and black holes - including their UV LF, the black mass function (BHMF) and the BH mass-stellar mass relations. The GW event rates predicted by this work are therefore benchmarked against all available high-$z$ galaxy and BH data. 

The cosmological parameters used in this work correspond to ($\Omega_{\rm m },\Omega_{\Lambda}, \Omega_{\rm b}, h, n_s, \sigma_8) = (0.3089,0.6911,0.049, 0.67, 0.96, 0.81)$, consistent with the latest results from the {\it Planck} collaboration \citep{planck2015}. We quote all quantities in comoving units unless stated otherwise and express all magnitudes in the standard AB system \citep{oke-gunn1983}.

The paper is organized as follows. In Section \ref{sec_model}, we detail our code for the galaxy-BH (co)-evolution, that we test against observations in Section \ref{testing_code}. In Section \ref{lisa}, we simulate {\it LISA's} performance in detecting our mock population of merging black holes and our results are summarised and discussed in Section \ref{conclusions}.

\section{The Theoretical model}
\label{sec_model}
This work is based on using the code {\it Delphi} ({\bf D}ark Matter and the {\bf e}mergence of ga{\bf l}axies in the e{\bf p}oc{\bf h} of re{\bf i}onization), introduced in previous papers including \citet{dayal2014a, dayal2015, dayal2017a, dayal2017b}. In brief, {\it Delphi} uses a binary merger tree approach to jointly track the build-up of dark matter halos, their baryonic component (both gas and stellar mass) and their star-formation driven spectra through cosmic time. We start by building merger trees for 550 $z=4$ galaxies, uniformly distributed in the halo mass range of $\log(M_h/ \Msun)=8-13.5$, up to $z=20$. Each $z=4$ halo is assigned a co-moving number density by matching to the $\der n / \der M_h$ value of the $z=4$ Sheth-Tormen halo mass function (HMF) and every progenitor halo is assigned the number density of its $z=4$ parent halo; we have confirmed that the resulting HMFs are compatible with the Sheth-Tormen HMF at all $z$. In terms of feedback, so far, this model has focused on modelling the impact of (TypeII) supernovae (SNII) and reionization feedback on the formation of early galaxies. In this work, we extend our model to include the impact of BH seeding, growth and feedback on early galaxy formation. The very first progenitors of every $z=4$ halo, that mark the start of its assembly (the so-called ``starting leaves"), are assigned an initial gas mass that scales with the halo mass according to the cosmological ratio such that $M_g = (\Omega_b/\Omega_m) M_h$. Depending on their mass and redshift, such starting leaves can also be seeded with a black hole as explained in Sec. \ref{bhseeds}. We start by calculating the star formation efficiency of a halo and the gas-mass remaining after SN feedback (Sec. \ref{sf_fb}). If a halo hosts a BH, a part of the gas left after star formation and SN feedback can be accreted onto the black hole and the impact of black hole feedback is included as detailed in Sec. \ref{bh_fb}. At each step, we include both the impact of smooth-accretion and mergers in assembling the halo, baryonic (gas and stellar mass) and BH mass as detailed in Secs. \ref{acc} and \ref{mergers}. In our endeavour to build a model with minimal free parameters, we limit ourselves to two and four {\it mass- and $z$-independent} free parameters related to star formation and BHs, respectively. 

\subsection{Seeding halos with black holes}
\label{bhseeds}
We explore the two formation channels that yield the lightest (Pop III) and the most massive (DCBH) black hole seeds. At variance with previous models for gravitational waves from BH seeds that included only one type of BH seeds \citep[except for the post-processed``mixed models" in][]{2011PhRvD..83d4036S} here we consider the possibility of more than one BH formation mechanism operating in the (early) Universe, as generally expected \citep[e.g.,][]{2016MNRAS.457.3356V}. These BH seeds are planted {\it in the starting leaves of any halo} as now detailed:

\begin{figure}
\center{\includegraphics[scale=0.475]{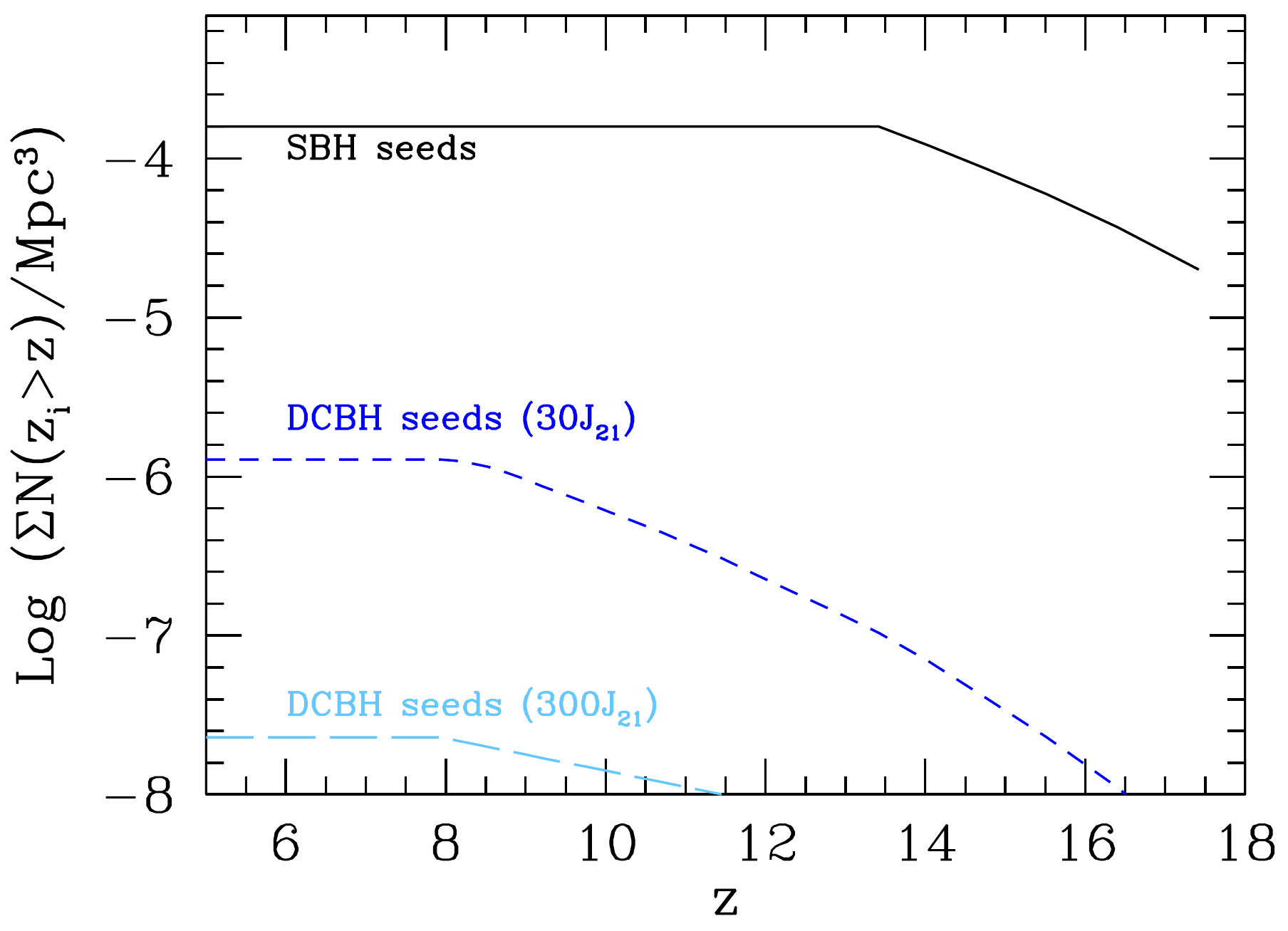}}
\caption{The cumulative redshift distribution of the number density of light (stellar black hole; solid black line) and heavy (DCBH) seeds in our model. For the latter we show the distribution of DCBH seeds using a value of $\alpha=30$ (short-dashed blue line) and $\alpha=300$ (long-dashed light-blue line).}
\label{fig_seeds} 
\end{figure}

(i) {\it Heavy seeds}: First postulated as massive ($10^{3-5}\Msun$) black hole seeds to explain the presence of SMBHs at early cosmic epochs \citep[e.g.][]{loeb1994, bromm2003}, DCBH formation models have been continually refined and developed over the past years \citep[e.g][]{begelman2006, begelman2008, regan2009, shang2010, johnson2012, latif2013, agarwal2014, dijkstra2014, ferrara2014, habouzit2016}. The current understanding from these works requires the following conditions to be met for a DCBH host: (i) the halo should have reached the atomic cooling threshold, with a virial temperature $T_{vir}\gsim 10^4 {\rm K}$, for the gas to be able to cool isothermally; (ii) the halo should be metal-free to prevent gas fragmentation; and (iii) the halo should be exposed to a high enough ``critical" Lyman-Werner (LW) background ($J_{crit} = \alpha J_{21}$). Here $\alpha>1$ is a free parameter and $J_{21}$ is the LW background expressed in units of $10^{-21} {\rm erg\, s^{-1}\, Hz^{-1} \, cm^{-2} \, sr^{-1}}$ (see e.g. \citealt{sugimura2014}). Interested readers are refereed to \citet{dayal2017b} for complete details on how DCBHs are seeded in high-$z$ halos. In brief, we start by making the reasonable assumption that the starting leaves of any halo are metal-free by virtue of never having accreted metal-enriched gas. Further, we use the stellar population synthesis code {\small Starburst99} \citep{leitherer1999} to calculate the LW ($11.2-13.6$ eV) luminosity of each galaxy based on its entire star formation history. This is used to calculate the mean LW emissivity at a given redshift, $\epsilon_{\rm LW}(z)$, by integrating over all galaxies present at that $z$. Accounting for fluctuations in the background, most likely around  galaxies and, from the biased (i.e. clustered) distribution of galaxies, we identify the probability of the starting leaves being irradiated by LW intensity above a critical threshold value as detailed in \citet{dayal2017b}; for the calculations in this work we explore values that range over an order of magnitude such that $\alpha = 30$ and $300$. 

The mass distribution of the seeds is uncertain and depends on the specific physical conditions at birth and on whether the intermediate state of a supermassive star is followed by a brief period of super-Eddington accretion onto the newly born BH (i.e. a quasistar phase). For example, the supermassive star mass depends on the strength of the LW radiation that illuminates the birth site \citep{2018arXiv180706337L,2018arXiv180708499A}. On the other hand, the existence of a quasistar phase and its outcome in term of BH seed mass depend on internal rotation and on mass loss in winds \citep{2011MNRAS.417.3035D,2016MNRAS.455....2F,2017MNRAS.464.2259F}. We are therefore left with an uncertain SMBH seed mass that can be bracketed by $10^{3}-10^{5}\Msun$ in halos below $10^9 \Msun$. Given the halo masses and LW radiation thresholds used in this paper \citep[cf. Fig.~5 in ][]{2018arXiv180706337L}, we randomly populate halos in the top half of the calculated probability range with a DCBH seed of mass ranging between $10^{3-4}\Msun$ ({\it ``light DCBH seeds"}): the number of halos populated with such DCBHs is calculated by matching this DCBH mass function to the probabilistic one (obtained by multiplying the mass function of DCBH hosts with the hosting probability). In order to check the dependence of our results on the DCBH seeds mass used, we also show the {\it LISA} event rates expected for seed masses higher by an order of magnitude, ranging between $10^{4-5}\msun$. The results of this {\it ``heavy DCBH seeds"} model are shown in Sec. \ref{lisa_results}.

(ii) {\it Light seeds}: Stellar BH seeds of mass $\sim 10^2\Msun$ can be created by the collapse of PopIII stars in minihalos with $M_h \sim 10^5 \Msun$. Making the reasonable assumption that halos collapsing from high ($\gsim 3.5$)-$\sigma$ fluctuations in the primordial density field are most likely to host such seeds \citep[e.g.][]{volonteri2003, barausse2012} results in the host halos being more massive than $M_h \gsim 10^{7.2}\Msun$ at $z \gsim 13$. However, given our halo mass range of $10^{8-13.5} \Msun$, starting leaves have to be assigned these seeds by hand. In this work, we start by populating the starting leaves of any halo, that fulfil the DCBH criterion detailed above, with seed DCBHs. The starting leaves of halos at $z \gsim 13$ that fulfil the light seed criterion, but do not contain a DCBH, are then populated with stellar BH seeds mass $M_{bh} = 150 \Msun$.
 
The initial seed distribution obtained with this formalism is shown in Fig. \ref{fig_seeds}. The cumulative number density of stellar BH seeds has a value of about $10^{-3.8}\, {\rm Mpc^{-3}}$ by $z \sim 4$. This clearly dominates over the cumulative number density of DCBH seeds which have a value of about $10^{-5.8}\, (10^{-7.6})\, {\rm Mpc^{-3}}$ by $z \sim 4$ for $\alpha = 30$ (300). Note that the models for ``light seeds" described above were inspired by early calculations that suggested that only one, very massive, PopIII star would form in a given halo, right at the center of the potential well \citep[e.g.,][]{2002Sci...295...93A}. More recent models favor a larger amount of fragmentation, leading to lighter and scattered PopIII remnants \citep{2014MNRAS.440.3778J,2018MNRAS.480.3762S} which are less suitable as SMBH seeds given the difficulty of both accreting material from their surroundings as well as finding the dynamical center of the galaxy/halo via dynamical friction. 

Therefore we also run test cases with only DCBHs as a lower limit to the presence of SMBH seeds in primeval galaxies. In terms of GW signatures, the results are similar to the reference case but selecting only mergers between heavy seeds. In terms of general BH population, models including only DCBHs with $\alpha = 300$ fail entirely in reproducing the observed active galactic nuclei (AGN) luminosity function at $z=5-6$, given their extremely low number densities; consider the long-dashed cyan line in Fig.~\ref{fig_seeds} which has a value of about $10^{-7.6}\, {\rm Mpc^{-3}}$ and compare this to an observed AGN number density of $\sim 10^{-5.5}-10^{-4.5} \, {\rm Mpc^{-3}}$ \citep{2018MNRAS.473.2378V,2018arXiv180709774K}. Models with $\alpha = 30$ are nearly compatible with the faint end of the observed luminosity function at $z=5$, but fail to produce enough AGN at $z=6$ unless the seed mass is above $10^4 \Msun$. In summary, both ``light" and ``heavy" seed models have uncertainties and problems associated with their formation and growth. Given this state-of-the-art in the field of BH formation, our model comprehensively explores the allowed parameter space, includes a realistic approach to the dependence of BH growth on the host mass, and presents a thorough comparison with observations in Sec.~\ref{testing_code} in order to explore the consequences of current uncertainties.

\subsection{Star formation and supernova feedback}
\label{sf_fb}
A newly-formed stellar population of mass $M_*(z)$ at redshift $z$ can impart the interstellar medium (ISM) with a total SNII energy $E_{SN}$ given by
\begin{equation}
E_{SN} = \fws E_{51} \nu M_*(z) \equiv \fws v_s^2 M_*(z).
\end{equation}
Here, each SNII is assumed to impart an (instantaneous) explosion energy of $E_{51}=10^{51}{\rm erg}$ to the ISM and $\nu = [134 \, {\rm \Msun}]^{-1}$ is the number of SNII per stellar mass formed for a Salpeter IMF between $0.1-100 \Msun$; we maintain this IMF through-out this work. The values of $E_{51}$ and $\nu$ yield $v_s= 611$ km s$^{-1}$. Finally, $\fws$ is the fraction of the SN explosion energy that couples to gas. 

For any given halo, the energy $E_{ej}$ required to unbind and eject the ISM gas not converted into stars can be expressed as
\begin{equation}
E_{ej}(z) = \frac{1}{2} [M_{gi}(z)-M_*(z)] v_e^2,
\end{equation}
where $M_{gi}(z)$ is the {\it initial} gas mass in the galaxy, prior to any star formation or BH accretion, at epoch $z$. Further, the escape velocity $v_e$ can be expressed in terms of the halo rotational velocity, $v_c$, as $v_e = \sqrt 2 v_c$.
 
We then define the {\it ejection efficiency}, $\fej$, as the fraction of gas that must be converted into stars to ``blow-away" the remaining gas from the galaxy (i.e.  $E_{ej} \le E_{SN} $). This can be calculated by imposing $E_{ej} = E_{SN} $ leading to 
\begin{equation}
\fej(z) = \frac{v_c^2(z)}{v_c^2(z) + \fws v_s^2}.
\label{fej}
\end{equation}
The {\it effective star formation efficiency} for any halo is then expressed as 
\begin{equation}
\feff =min[\fs,\fej],
\end{equation}
where $\fs$ is a free parameter representing the maximum instantaneous star formation efficiency - this parameter is fixed by matching to the bright-end of the observed LBG UV LF as explained in Sec. \ref{uvlf}. 
 
In this formalism, the newly formed stellar mass formed at $z$ can be expressed as
\begin{equation}
M_*(z) = M_{gi}(z) \feff.
\end{equation}

In the spirit of maintaining simplicity, we assume that every stellar population has a fixed metallicity of $0.05 \Zsun$ and each newly-formed stellar population has an age of $2 \, \myr$. Using these parameters with the population synthesis code {\small STARBURST99} \citep{leitherer1999}, the rest-frame UV luminosity (between $1250$ and $1500$\AA) from a newly-formed stellar mass can be expressed as
\begin{equation}
\luvs = 10^{33.077} \bigg(\frac{M_{*}}{\Msun} \bigg) \,\, {\rm erg\, s}^{-1} {\rm \AA}^{-1}.
\label{lumnew}
\end{equation}

This star-formation episode must then result in a certain amount of gas, $\mges(z)$, being ejected from the galaxy at the given $z$-step. The value of $\mges(z)$ depends on whether $\feff = f_*$ or $\feff = \fej$: while the galaxy is an ``efficient star-former" in the former case, that can support new stellar mass being formed without losing much of its gas, the latter case is true for a ``feedback-limited" system that loses all of its ISM gas after star formation. Mathematically, $\mges$ can be calculated as
\begin{equation}
\mges(z) = [M_{gi}(z) - M_{*}(z)] \frac{\feff}{\fej}.
\label{mej}
\end{equation}
The final gas mass, $\mgfs(z)$, remaining in the galaxy at that redshift-step, after star formation and SN feedback, can then be expressed as
\begin{equation}
\mgfs(z) = [M_{gi}(z) - M_*(z)] \bigg[1-\frac{\feff}{\fej}\bigg].
\label{mf}
\end{equation}

\begin{table*}
\centering
\caption {Values of the free parameters used for the models indicated in column 1. See text in sec. \ref{sec_model} for details. }
\centerlast
\begin{tabular}{ |c|c|c|c|c|c|c|c|c|c|}
 \hline
  Model &  $\alpha$ &  $UVB$ & $\fs$ & $\fws$ & $\fwb$ & $\faccb$ & $\fed (M_h < \mcritb)$ & $\fed (M_h > \mcritb)$ & $\epsilon_r$  \\
   \hline
  ins1 & 30 & No & 0.02 & 0.1 & 0.003 & $5.5 \times 10^{-4}$  & $7.5 \times 10^{-5}$ & 1 & 0.1 \\
  ins2 & 300 & No & 0.02 & 0.1 & 0.003 & $5.5 \times 10^{-4}$  & $7.5 \times 10^{-5}$ &  1 & 0.1 \\
  ins3 & 30 & Yes & 0.02 & 0.1 & 0.003 & $5.5 \times 10^{-4}$  & $7.5 \times 10^{-5}$ & 1 & 0.1  \\
  ins4 & 300 & Yes & 0.02 & 0.1 & 0.003 & $5.5 \times 10^{-4}$  & $7.5 \times 10^{-5}$ & 1 & 0.1  \\
  tdf1 & 30 & No & 0.02 & 0.1 & 0.003 & $5.5 \times 10^{-4}$  & $7.5 \times 10^{-5}$ & 1 & 0.1  \\
  tdf2 & 300 & No & 0.02 & 0.1 & 0.003 & $5.5 \times 10^{-4}$  & $7.5 \times 10^{-5}$ &  1 & 0.1  \\
  tdf3 & 30 & Yes & 0.02 & 0.1 & 0.003 & $5.5 \times 10^{-4}$  & $7.5 \times 10^{-5}$ & 1 & 0.1  \\
  tdf4 & 300 & Yes& 0.02 & 0.1 & 0.003 & $5.5 \times 10^{-4}$  & $7.5 \times 10^{-5}$ & 1 & 0.1  \\
  \hline
 \end{tabular}
 \label{table1}
\end{table*} 
\subsection{Black hole growth and feedback}
\label{bh_fb}
Once seeded, BHs can grow via accretion and mergers. We discuss BH growth via accretion in this section and the merger-driven growth is deferred to Sec. \ref{mergers} that follows. At any given redshift, the Eddington mass accretion rate, $\dot M_{ed}$, for a BH of mass $\mbh$ can be calculated as
\begin{equation}
\dot M_{ed}(z) = \frac{4 \pi G \mbh (z) m_p}{\sigma_T \epsilon_r c},
\end{equation}
where $G$ is the gravitational constant, $m_p$ is the proton mass, $\sigma_T$ is the Thomson scattering optical depth, $\epsilon_r$ is the BH radiative efficiency and $c$ is the speed of light. Given our merger tree time-step of $\Delta t = 20 {\rm Myr}$, the total mass that can be accreted at the Eddington rate in one time-step is $\med (z) = (1-\epsilon_r) \dot M_{ed}(z) \times \Delta t$. 

Further, the gas mass accreted by the BH in a given time step, $\maccb(z)$, is calculated as:
\begin{equation}
    \maccb(z)=
    \begin{cases}
      \fed \med(z), & {\rm if}\ x_r\faccb  \mgfs (z)> \fed \med(z) \\
      x_r \faccb  \mgfs(z), & {\rm if}\ x_r\faccb \mgfs(z)< \fed \med(z)
    \end{cases}
  \end{equation}
where $x_r = (1-\epsilon_r)$. Further, $\faccb$ is the fraction of the available gas mass left, after star formation and supernova feedback, that can be accreted by the BH and $\fed$, a free-parameter, is the fractional Eddington rate of accretion. Using this formalism, the BH accretes either at a fraction of the Eddington rate or a fraction of the available gas mass, whichever is lower, i.e., $\maccb(z) = min[x_r \faccb  \mgfs(z), \fed \med(z)]$. Matching to the AGN UV LF (Sec. \ref{uvlf}) requires $\fed = 7.5 \times 10^{-5}$ below a critical halo mass of $\mcritb = 10^{11.25}[\Omega_m(1+z)^3 +\Omega_\Lambda ]^{-0.125}$ \citep[see also][]{bower2017} and $\fed = 1$ above this mass at any redshift. This accretion will yield a BH feedback energy of 
\begin{equation}
E_{bh} = \fwb \epsilon_r \maccb(z) c^2,
\end{equation}
 where $\fwb$ is the efficiency of BH feedback coupling to the gas. 
 
 The BH feedback required to eject the left-over gas (after accretion) can be expressed as
 \begin{equation}
 E_{bh}^{ej}(z) = \frac{1}{2}[\mgfs(z) - \maccb(z)] v_e^2.
 \end{equation}
The {\it effective black hole feedback} is therefore taken to be the minimum between the energy required to eject all the gas up to the maximum value such that $E_{bh}^{eff} = min[E_{bh}, E_{bh}^{ej}]$. The final gas mass left in the halo, that can be carried over for mergers, after BH accretion and feedback can then be calculated as
 \begin{equation}
 \mgfb(z) = [\mgfs(z) - \maccb(z)] \bigg[1-\frac{E_{bh}^{eff}}{E_{bh}^{ej}}\bigg]. 
 \end{equation}
 
Using the above formalism, the total luminosity produced by the black hole can be expressed as
\begin{equation}
L_{bh} = \frac{\epsilon_r \maccb(z) c^2}{\Delta t} \, [L_\odot].
\end{equation}
This is converted into the B-band luminosity using the results of \citet{marconi2004} where ${\rm log(L_{bh}/\nu_B L_{\nu B} )} = {\rm 0.80 -0.067({\rm log}\, L_{bh}-12)} + {\rm 0.017 ({\rm log}\,  L_{bh}-12)^2} - {\rm 0.0023 ({\rm log} \, L_{bh}-12)^3}$. Finally, we use $L_\nu \propto \nu^{-0.44}$ to convert this B-band luminosity into the black hole UV luminosity $\luvb$. In order to build the AGN UV LF, we also account for AGN obscuration by multiplying the number density of BHs of a given luminosity by the correction factors proposed in \citet{ueda2014}.

\subsection{Smooth-accretion from the intergalactic medium}
\label{acc}
The merger of halos is accompanied by ``smooth-accretion" of dark matter from the intergalactic medium (IGM). In the analytic merger tree, this smoothly-accreted dark matter mass is calculated as
\begin{equation}
\mdmsa(z) = M_h(z) - \sum_{i=1}^{N} M_h (z+\Delta z),
\end{equation} 
where $M_h(z)$ is the halo mass at $z$ and the second term on the RHS denotes the sum of the halo masses of all its progenitors at the previous redshift step; $N=2$ in the case of the binary merger tree used in our study.

We make the reasonable assumption that the smooth accretion of dark matter is accompanied by the accretion of a cosmological ratio of gas from the IGM such that the smoothly-accreted gas mass $\mgsa$ at $z$ can be written as
\begin{equation}
\mgsa(z) = \frac{\Omega_b}{\Omega_m} \mdmsa(z).
\end{equation}
\subsection{Merging galaxies and black holes}
\label{mergers}

As galaxies merge, in addition to dark matter, they bring in stellar and gas mass with the latter depending on the star formation and BH accretion efficiencies of the progenitor halos. As detailed in Secs. \ref{sf_fb} and \ref{bh_fb}, galaxies forming stars and/or accreting at the limit of BH feedback will only bring stellar mass into their successors resulting in ``dry mergers". On the other hand, halos bringing in both stellar and gas mass result in ``wet mergers". Mergers of galaxies result in a summation of their dark matter, baryonic and BH masses and total UV luminosities such that
\begin{eqnarray}
M_{dm}(z) & = & \sum_{i=1}^N M_h(z+\Delta z) + \mdmsa(z)\\
M_{g}(z) & = & \sum_{i=1}^N \mgfb(z+\Delta z) + \mgsa(z) \\
M_{*, tot}(z) & = & M_{*}(z) + \sum_{i=1}^N M_{*}(z+\Delta z) \\
M_{bh, tot}(z) & = & M_{bh}(z) + \sum_{i=1}^N M_{bh}(z+\Delta z) \\
\luvtot(z) & = & \luvs(z) + \luvb(z) +  \sum \luvs(z+\Delta z).
\end{eqnarray}
Using {\small STARBURST99}, we find that the UV luminosity for a burst of stars 
(normalized to a mass of $1 \Msun$ and metallicity $0.05\, \Zsun$) decreases with time as 
\begin{equation}
\log \bigg(\frac{\luvs(t)}{{\rm erg\, s}^{-1} {\rm \AA}^{-1}} \bigg) = 33.0771 - 1.33 \log (t/t_0)  + 0.462,
\label{luv}
\end{equation}
where $t$ is the age of the stellar population (in yr) at $z$ and $\log (t_0/{\rm yr}) = 6.301$. Finally, we make the limiting assumption that BH luminosity decays away within the time-step of $20\,  \myrs$ i.e. BH luminosity is only relevant in the redshift step in which the black hole accretes.

We also explore two scenarios for the {\it timescales} of galaxy mergers: {\it in the first}, halos mergers are accompanied by the mergers of galaxies and their constituent BHs. This {\it instantaneous merging} scenario sets the {\it upper limit} for the BH merger rate. {\it In the second scenario}, we include the fact that galaxies (and their BHs) merge after a ``merging" timescale which can be calculated as \citep{lacey-cole1993}
\begin{equation}
\tau = f_{df} \Theta_{orbit} \tau_{dyn} \frac{M_{host}}{M_{sat}} \frac{0.3722}{ln(M_{host}/M_{sat} )},
\end{equation}
where $M_{host}$ is the mass of the host including all the satellites, $M_{sat}$ is the mass of the merging satellite, $\tau_{dyn}$ represents the  dynamical timescale and $f_{df}$ represents the efficiency of tidal stripping; $f_{df}>1$ if tidal stripping is very efficient. For this work, we use $f_{df}=1$. Further, 
\begin{equation}
\Theta_{orbit} = \bigg(\frac{J}{J_c}\bigg)^{0.78} \bigg(\frac{r_c}{R_{vir}}\bigg)^2,
\end{equation}
where $J$ is the satellite's specific angular momentum and $J_c$ that of a satellite
carrying the same energy and orbiting on a circular orbit. The last term represents the ratio between the circular radius (the radius of a circular orbit with the same energy) and the virial radius of the host. $\Theta_{orbit}$ is well modelled by a log-normal distribution such that ${\rm log} (\Theta_{orbit}) = -0.14 \pm 0.26$ \citep{cole2000}; we randomly sample values from this distribution for each merger. Finally, the dynamical timescale can be calculated as $\tau_{dyn} = \pi R_{vir}(z) V_{vir}(z)^{-1} = 0.1 \pi t_H(z)$ where $R_{vir}$ and $V_{vir}$ are the virial radius and velocity at $z$. We also make the limiting assumption that satellite galaxies, waiting to merge, neither form stars nor have any BH accretion of gas. 

We show results from both scenarios in this work in order to bracket the uncertainty on our understanding of galaxy merger timescales. In reality, however, the second scenario still gives a lower limit to the merger time of BH binaries: additional delays are to be expected once the binary forms after the dynamical friction timescale. The binary has to harden further to reach the separation where gravitational wave emission becomes the dominant source of energy and momentum losses to bring the binary to coalescence and merge within the Hubble time. This additional delay is largely unconstrained, and can range from tens to billions of years \citep[for a review see, e.g.,][]{barack2018}. Additionally, for low-mass BHs dynamical friction could be ineffective during the galaxy merger phase since the deceleration of dynamical friction is proportional to the infalling black hole mass resulting in the timescale for orbital decay being inversely proportional to mass.

\subsection{The impact of reionization}
\label{reio}
In this work, we also include the effects of the  Ultra-violet background (UVB) created during reionization which, by heating the ionized IGM to $T \sim 10^4$ K,  can have an impact on the baryonic content of low-mass halos \citep[for a recent review see, e.g.,][] {dayal2018}. Given that self-consistently calculating the impact of the UVB on galaxy formation remains an unsolved problem, we consider two scenarios: {\it the first} is in which UVB has no impact on the baryonic content of any halo. {\it The second} scenario is one considering maximal UVB feedback in which the gas mass is completely photo-evaporated for all halos below a characteristic virial velocity of $V_{vir} = 40 \, \kms$. Critically, whilst leaving the {\it number} of mergers unchanged, limiting the gas mass available for accretion onto BHs, the latter scenario results in a {\it lower limit} on the mass of the merged BH. 

\begin{figure*}
\center{\includegraphics[scale=1.01]{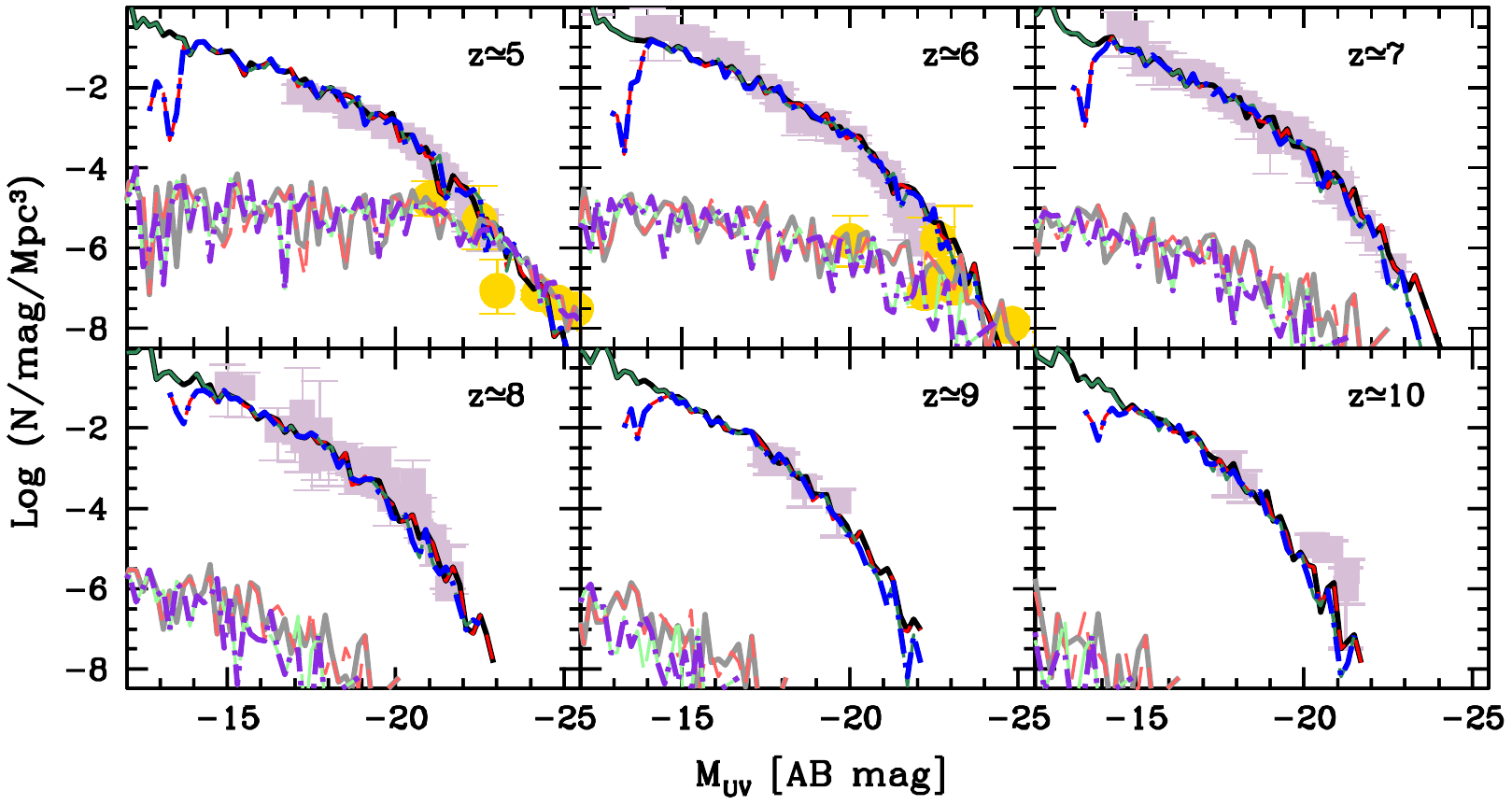}}
\caption{The UV LF from $z \simeq 5-10$ as marked in the panels. In each panel, the violet points show the available LBG data collected both using space- and ground-based observatories at: (a) $z\simeq 5$ \citep{bouwens2007, mclure2009}; (b) $z\simeq 6$ \citep{mclure2009, bouwens2015, livermore2017}; (c) $z \simeq 7$ \citep{bouwens2010a, mclure2010, castellano2010, mclure2013, bowler2014a, livermore2017, atek2015}; (d) $z \simeq 8$ \citep{bouwens2010a, mclure2010, bradley2012, mclure2013, livermore2017, atek2015}; (e) $z \simeq 9$ \citep{mclure2013, oesch2013}; and (e) $z \simeq 10$ \citep{bouwens2014, oesch2014}. In each panel, the yellow points show the AGN data collected at: $z \sim 5$ \citep{mcgreer2013, parsa2018} and $z \sim 6$ \citep{willott2010,kashikawa2015,parsa2018,jiang2016}. In each panel, lines show model UV LFs for galaxies and black holes for the following models summarised in Table 1, that bracket the range of UV LFs allowed in the presence/absence of a UVB and for both instantaneous and delayed (by a merging timescale) merger: {\it ins1} ({\it galaxies} solid black line; {\it BH} solid gray line), {\it ins4} ({\it galaxies} short dashed red line; {\it BH} short dashed light-red line), {\it tdf1} ({\it galaxies} long dashed green line; {\it BH} long-dashed light green line) and {\it tdf4} ({\it galaxies} dot-dashed blue line; {it BH} dot-dashed purple line). }
\label{fig_uvlf} 
\end{figure*}

We run the above model for 8 different scenarios (detailed in Table \ref{table1}) that explore 3 key physical effects: {\it (i)} the impact of the LW background amplitude in calculating DCBH hosts (Sec. \ref{bhseeds}); {\it (ii)} the impact of instantaneous versus delayed mergers of galaxies and BHs (Sec. \ref{mergers}); and {\it(iii)} the impact of UV feedback on early galaxies and BHs (Sec. \ref{reio}). In what follows, the model {\it ins1}, that assumes the lowest LW amplitude for DCBHs, instantaneous galaxy mergers and no UV feedback, is denoted as the {\it fiducial} model and provides the upper limit on our results. On the other hand, with the highest LW amplitude for DCBHs, delayed galaxy (and BH) mergers and maximal UV feedback, the model {\it tdf4} provides the lower limit to our results.

\section{Comparing theoretical galaxy and black hole properties to observations}
\label{testing_code}
Now that the theoretical model has been established, we compare model results to a number of observational data sets, for both galaxies and AGN, as detailed in what follows. In terms of galaxies, we use the data sets accumulated for  high-$z$ LBGs, detected through a drop in luminosity blue-ward of the Lyman limit at $912$\AA. The past few years have seen an enormous increase in LBG data due to a combination of state-of-the-art instruments such as (the Wide Field Came 3 onboard) the Hubble Space Telescope ({\it HST}) as well as refined selection techniques \citep[e.g.][]{steidel1999}. As for AGN, a number of surveys, including the Sloan Digital Sky Survey (SDSS) and the Canadian-French high-z quasar surveys, and observations with the Subaru telescope, have yielded a statistical sample of AGN/QSO candidates at redshifts as high as $z \simeq 6$. In what follows we compare 4 models that bracket the physically plausible range explored in this work ({\it ins1, ins4, tdf1 and tdf4}), detailed in Table \ref{table1}, with a number of data-sets, including the UV LFs, the stellar mass density, the black hole mass function and the black hole-stellar mass relation.

We note that, given their low number densities, both the ``light" and ``heavy" DCBH seeding cases yield very similar results for all the observational data-sets discussed. For this reason, we limit our results to the ``light DCBH seed" case in this section. 

\subsection{The observed UV LF for star formation and black holes}
\label{uvlf}
The observed UV LF (number density of galaxies as a function of the absolute magnitude) and its redshift evolution offer one of the most robust tests of theoretical models of galaxy formation. We start by calculating the UV magnitudes, separately for star formation and AGN activity, for each theoretical galaxy and computing the associated UV LFs, as shown in Fig. \ref{fig_uvlf}. We start by discussing the LBG UV LF: firstly, matching to the bright end of the evolving UV LF requires a maximum star formation efficiency value of $f_* \simeq 2\%$. Secondly, we find that the fiducial model ({\it ins1}) is in excellent agreement with available LBG observations, ranging between $-22 \lsim \muv \lsim -13$, at all $z \sim 5-10$ as already shown in our previous works \citep[e.g.][]{dayal2014a}. The inclusion of a delay in galaxy mergers ({\it tdf1}) has no sensible impact on the faint-end of the UV LF - this is due to the fact that the progenitors of these low-mass halos are SN feedback limited and hence do not bring in any gas whilst merging (dry mergers) as already pointed out previously \citep{dayal2014a}. On the other hand, the delay in galaxy mergers leads to an increasing reduction in the gas masses of higher-mass halos whose progenitors are not SN feedback limited and bring in gas in mergers (wet mergers), leading to a slightly steeper bright end. Finally, including the (maximal) impact of reionization feedback ({\it ins4} and {\it tdf4}), that photo-evaporates the baryonic content of all galaxies with $V_{vir} \lsim 40 \, \kms$, only affects the faint-end of the UV LF and leads to a cut-off at brighter magnitudes ($\muv \sim -14$ to $-15$) as compared to the continued rise excluding this effect (e.g. in models {\it ins1} and {\it tdf1}). In this work, models {\it ins1} and {\it tdf4}, therefore, bracket the plausible UV LF range. However, it must be cautioned that the theoretical LBG UV LF has, so far, ignored the impact of dust enrichment which is expected to have a relevant effect in decreasing the luminosities at the bright end. 

Focusing on the AGN UV LF, the black hole powered UV LFs for all four models discussed above are found to be in excellent agreement with all available AGN data at $z \sim 5$ and $6$ as shown in Fig. \ref{fig_uvlf}. We start by noting that given the large masses ($M_h \gsim 10^{11.5}\msun$) associated with AGN/QSO host halos, the black hole UV LF is only relevant at $\muv \lsim -21$, corresponding to number densities $\lsim 10^{-5} [{\rm dex^{-1} Mpc^{-3}}]$ at $z \sim 5$ and $6$. These results are in qualitative agreement with those of \cite{ono2018} who find 100\% of the UV luminosity to come solely from stars for galaxies with $\muv \gsim -23$ to $-24$. However, given that the AGN number densities are suppressed due to obscuration (see Sec. \ref{bh_fb}), calculating the fraction of galaxies dominated by AGN requires a more thorough examination which we defer to a future work. Finally, we note that the contribution of BH-powered luminosity could be one explanation for observed UV LFs that are shallower than the exponentially declining Schechter function at these high-$z$ \citep[e.g.][]{ono2018}.

We find that the AGN UV LF is extremely similar for heavy black hole seeds with $\alpha$ varying over an order of magnitude (for 30 to 300) for the four models discussed above. This is probably to be expected given the extremely low number of heavy black hole seeds as compared to the number of light black hole seeds as pointed out in Sec. \ref{bhseeds}; the latter therefore clearly dominate the UV LF. As for the merger timescales, including a delay in the mergers of galaxies (and black holes) results in a smaller black hole growth. This is reflected in a lower final black hole mass in a given halo (also see Sec. \ref{bhmf} that follows). However, this only leads to minor changes in the UV LF which are indistinguishable within the scatter shown by the four models considered here. Further, given the large masses of AGN hosts, reionization feedback has no relevant effect on the AGN UV LF. Finally, looking at the redshift evolution of the AGN UV LF, we find that it shows a sharper redshift evolution compared to the star-formation powered UV LF given the increasing paucity of their high-mass hosts. To quantify this effect, let us focus on a magnitude of $\muv =-20$: while the star formation driven UV LF only evolves by a factor of 3 between $z \sim 5$ and $7$, the AGN UV LF (negatively) evolves by roughly three orders of magnitude over the same redshift range. 

\subsection{The LBG stellar mass density (SMD)}
\label{smd}

\begin{figure}
\center{\includegraphics[scale=0.475]{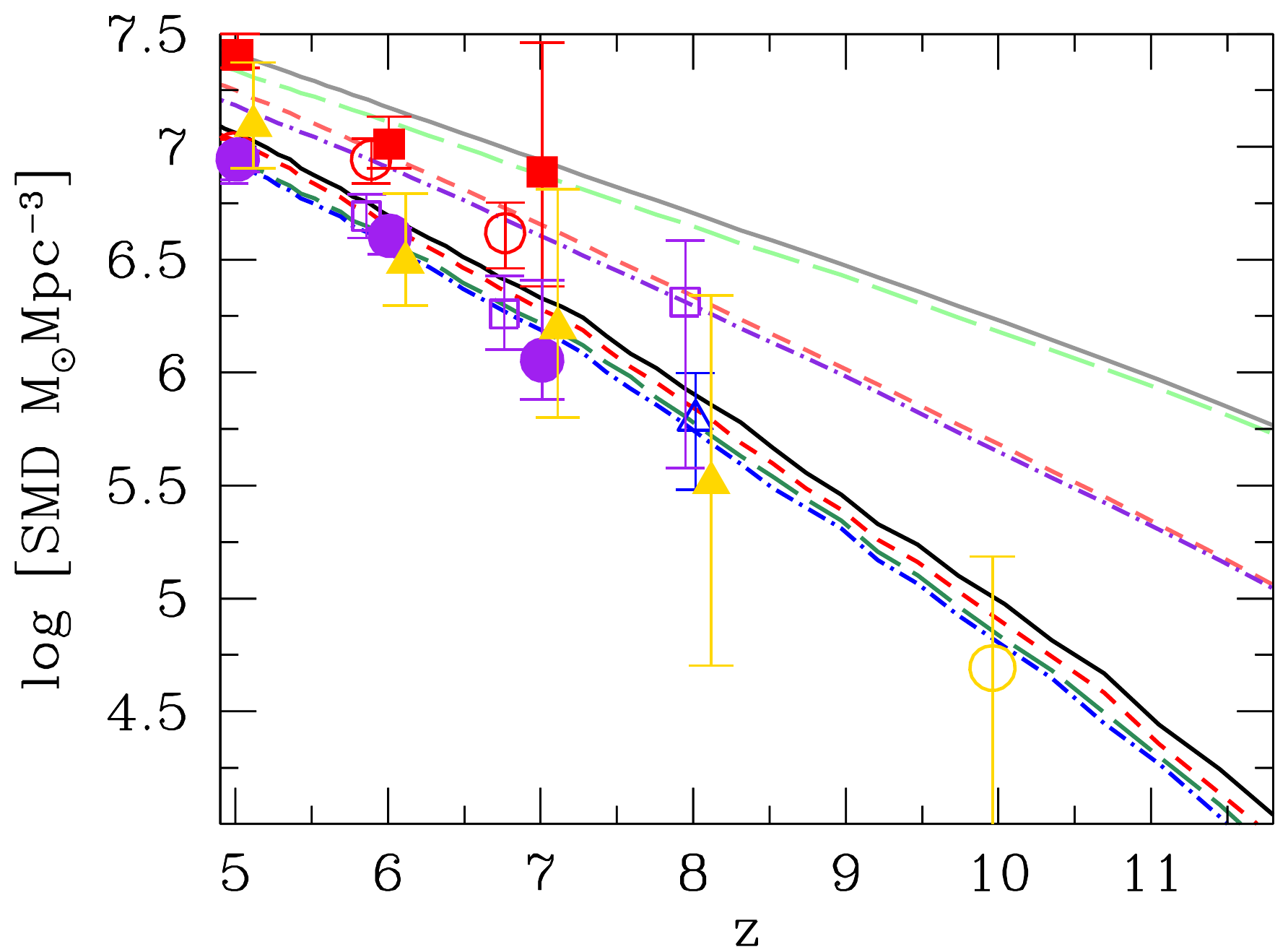}}
\caption{The LBG stellar mass density (SMD) as a function of redshift. Points show the observational data collected by: \citet[][red empty circles]{gonzalez2011}, \citet[][blue empty triangles]{labbe2013}, \citet[][purple empty squares]{stark2013}, \citet[][yellow empty circles]{oesch2014}, \citet[][red filled squares]{duncan2014}, \citet[][purple filled circles]{grazian2015} and \citet[][yellow filled triangles]{song2016}. We show results for galaxies with $\muv <-17.7$ which can be directly compared to observational data points for the following models shown in Table 1: {\it ins1} (solid black line), {\it ins4} (dot-dashed red line), {\it tdf1} (long dashed green line) and {\it tdf4} (dot-dashed blue line). We also show results for the total SMD obtained by summing over all galaxies at a specific $z$ for the same models noted above: {\it ins1} (solid gray line), {\it ins4} (dot-dashed light red line), {\it tdf1} (long dashed light green line) and {\it tdf4} (dot-dashed purple line). }
\label{fig_smd} 
\end{figure}

We now compare the theoretical SMD to that observationally inferred for LBGs. We start by comparing to observed LBGs with $\muv \lsim -17.7$ as shown in Fig. \ref{fig_smd}. As seen, while all four models ({\it ins1, ins4, tdf1, tdf4}) are in excellent agreement with the data they are offset in normalisation from each-other whilst following very similar slopes such that $SMD \propto (1+z)^{0.42}$. As might be expected, model {\it ins1} provides the upper limit to the SMD results for observed galaxies. Including the effects of delayed galaxy mergers ({\it 
tdf1}) results in a small decrease in the SMD values by about 0.1 dex. However, assuming instantaneous mergers whilst including maximal UVB suppression ({\it ins4}) only results in a SMD that is different from the fiducial case by a negligible 0.03 dex. These results clearly imply that a delay in the merger timescales is more important than the effect of a UVB for these high mass systems. Finally, the lower limit to the SMD results is provided by model {\it tdf4} that is about 0.13 dex lower than the fiducial results. These slight changes in the SMD normalisation shows that most of the stellar mass is assembled in massive progenitors \citep[see also][]{dayal2013} with low-mass progenitors - that either merge after a dynamical timescale ({\it tdf1}), are reionization suppressed ({\it ins4}) or include both these effects ({\it tdf4}) - contributing only a few percent to the stellar mass for observed galaxies.
 
 \begin{figure}
\center{\includegraphics[scale=0.475]{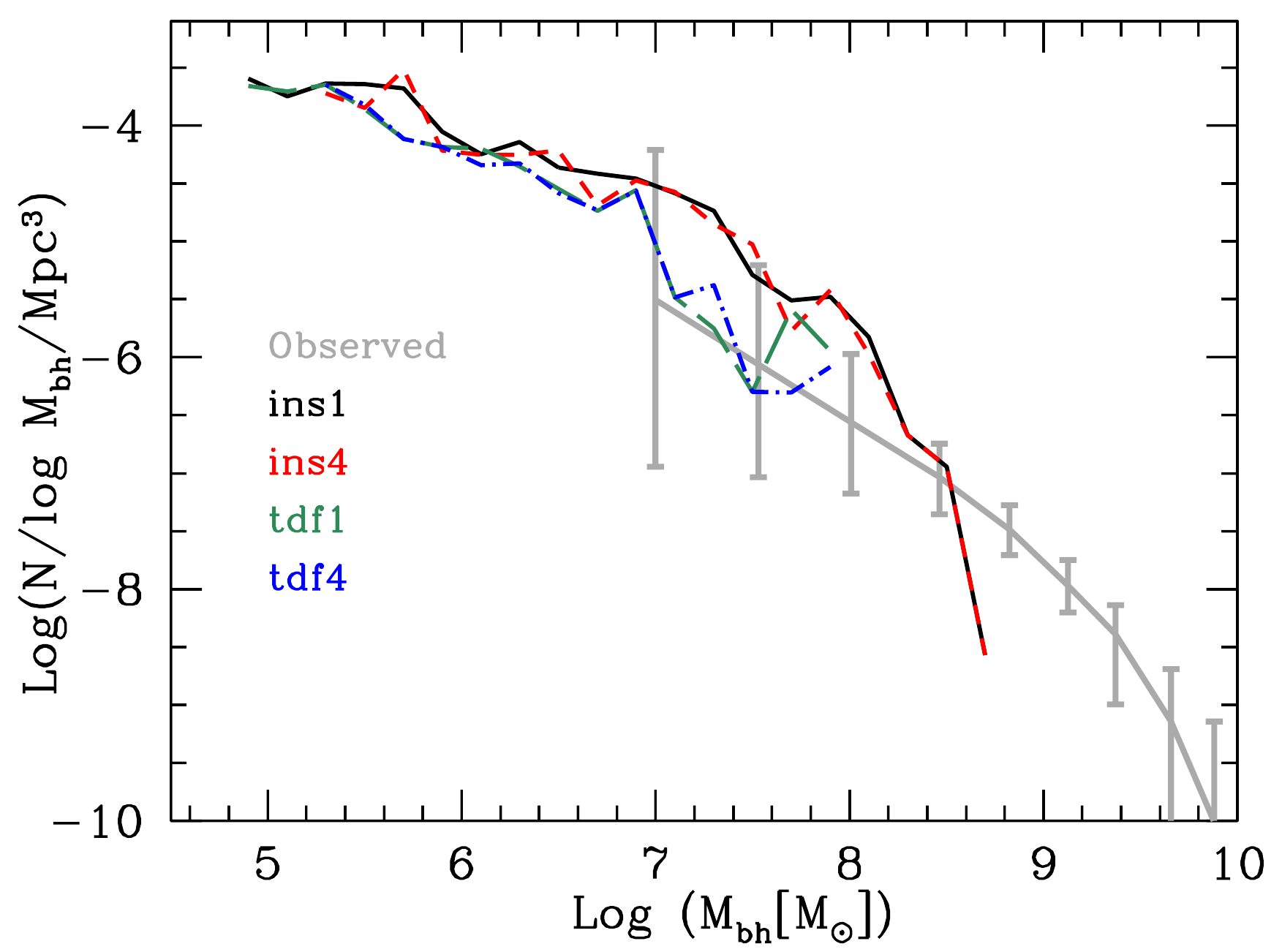}}
\caption{The black hole mass function (BHMF) at $z\simeq 6$. We compare observational results (gray line with error bars) from \citet{willott2010} to those from our models bracketing the plausible physical range: {\it ins1} (solid black line), {\it ins4} (short-dashed red line), {\it tdf1} (long-dashed green line) and {\it tdf4} (dot-dashed blue line). As shown, the shape of the BHMF is independent of the inclusion of UV feedback and the merger timescales used. However, the final BH masses are naturally lower when including a delay in the merger timescales as opposed to instantaneous mergers.  }
\label{fig_bhmf} 
\end{figure}

On the other hand, the impact of reionization feedback and a delay in the merging timescale are much more dramatic when considering the entire galaxy population (without any limiting magnitudes used). In this case, the fiducial model, {\it ins1}, shows a slope that evolves with redshift as $SMD \propto (1+z)^{0.24}$. Given that in this case the SMD is dominated by the contribution from low-mass halos, the situation flips as compared to that discussed above: the merger timescale has a negligible effect on the SMD of all galaxies and shows essentially the same amplitude and slope as the fiducial case. However, the UVB suppression of the gas mass of low-mass halos results in both a decrease in the amplitude (by about 0.23 dex) and a, more dramatic, steepening of the SMD slope such that $SMD \propto (1+z)^{0.31}$ for models {\it ins4} and {\it tdf4}. 
 
\begin{figure}
\center{\includegraphics[scale=0.475]{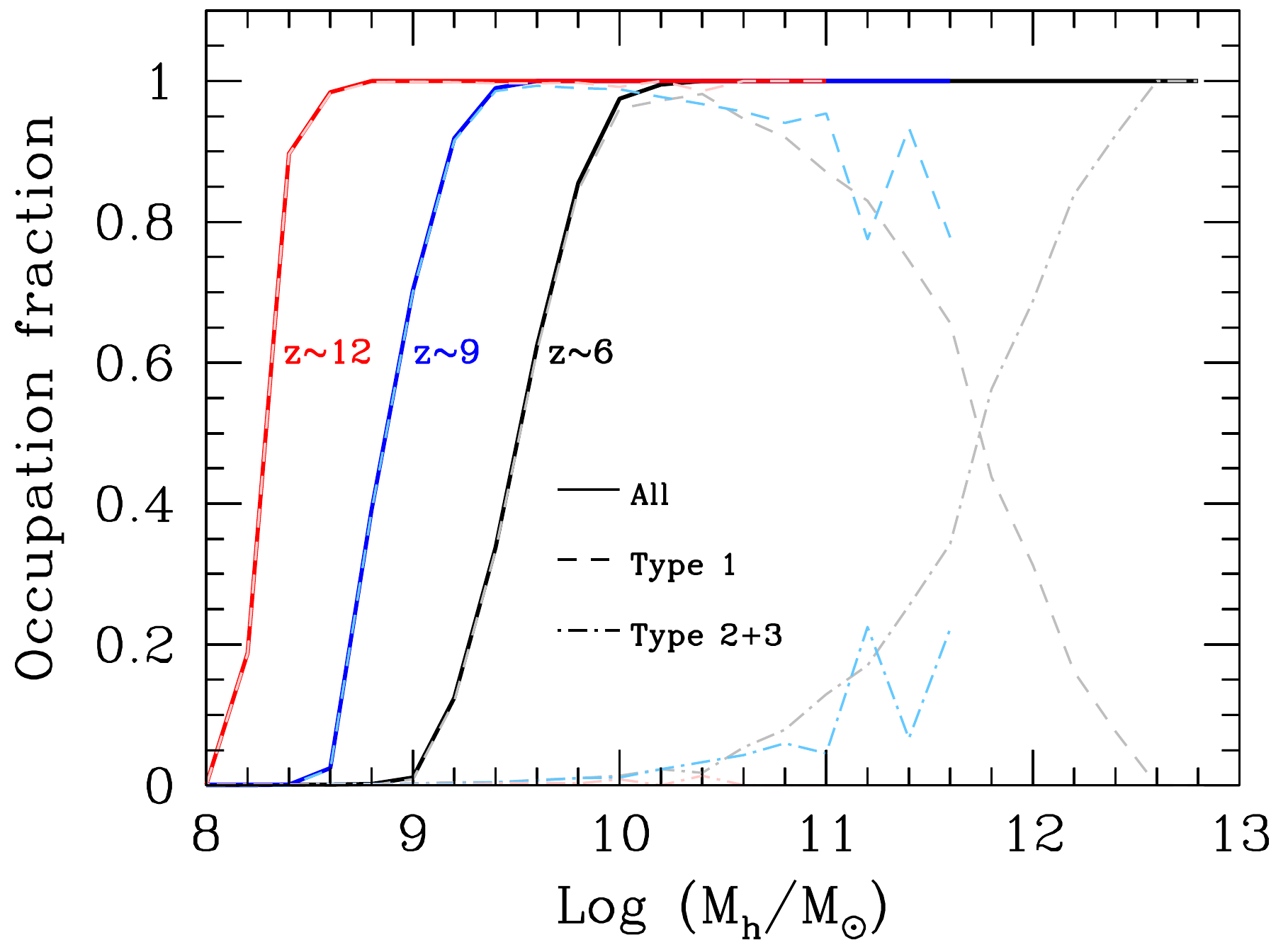}}
\caption{The black hole occupation fraction as a function of halo mass for $z \sim 6$ (black lines), $z \sim 9$ (blue lines) and $z \sim 12$ (red lines) as marked. The solid lines show the occupation fraction for all black holes; the short-dashed and dot-dashed lines show results for Type 1 (stellar black holes) and Type 2+3 (DCBHs), respectively.  }
\label{fig_occu} 
\end{figure}
 
\subsection{The black hole mass function and occupation fraction}
\label{bhmf}
We now discuss the black hole mass function (BHMF) which expresses the number density of black holes as a function of their mass, the results of which at $z \simeq 6$ are shown in Fig. \ref{fig_bhmf}. As expected, the number density of black holes increase with decreasing BH mass as shown in the Figure. The observed BHMF at $z \sim 6$ extends from $M_{bh} \sim 10^{7-10}\Msun$. Our theoretical results for all four models discussed above are in good agreement with the data within error bars as seen in the same figure. Naturally, the fiducial model ({\it ins1}), extending from $M_{bh} \sim 10^{4.8-8.8}\Msun$, yields the upper limit to the BHMF. Including a delay in the merger times for black holes ({\it tdf1}) leads to a decrease in the maximum mass attained by the black holes ($M_{max} \sim 10^8\Msun$) showing that gas brought in by merging progenitors halos has a significant contribution to the growth of these high-mass systems. On the other hand, reionization feedback alone ({\it ins4}) has a negligible effect on the growth of high-mass halos (as discussed in Sec. \ref{smd} above), yielding a BHMF in close agreement with the fiducial one. Finally, the model {\it tdf4}, including both the impact of delayed mergers and the UVB, yields results quite similar to {\it tdf1} and, provides the lower limit to the BHMF.  We recall that our model is not aimed at (re)producing rare luminous quasars powered by very massive BH \citep[see][and references therein for models focused on the most massive halos and BHs]{2016MNRAS.457.3356V,2016MNRAS.458.3047P} but at the bulk of the population of massive BHs. It should therefore not be surprising that the BH mass function does not extend to the highest BH masses observed.

\begin{figure*}
\center{\includegraphics[scale=1.01]{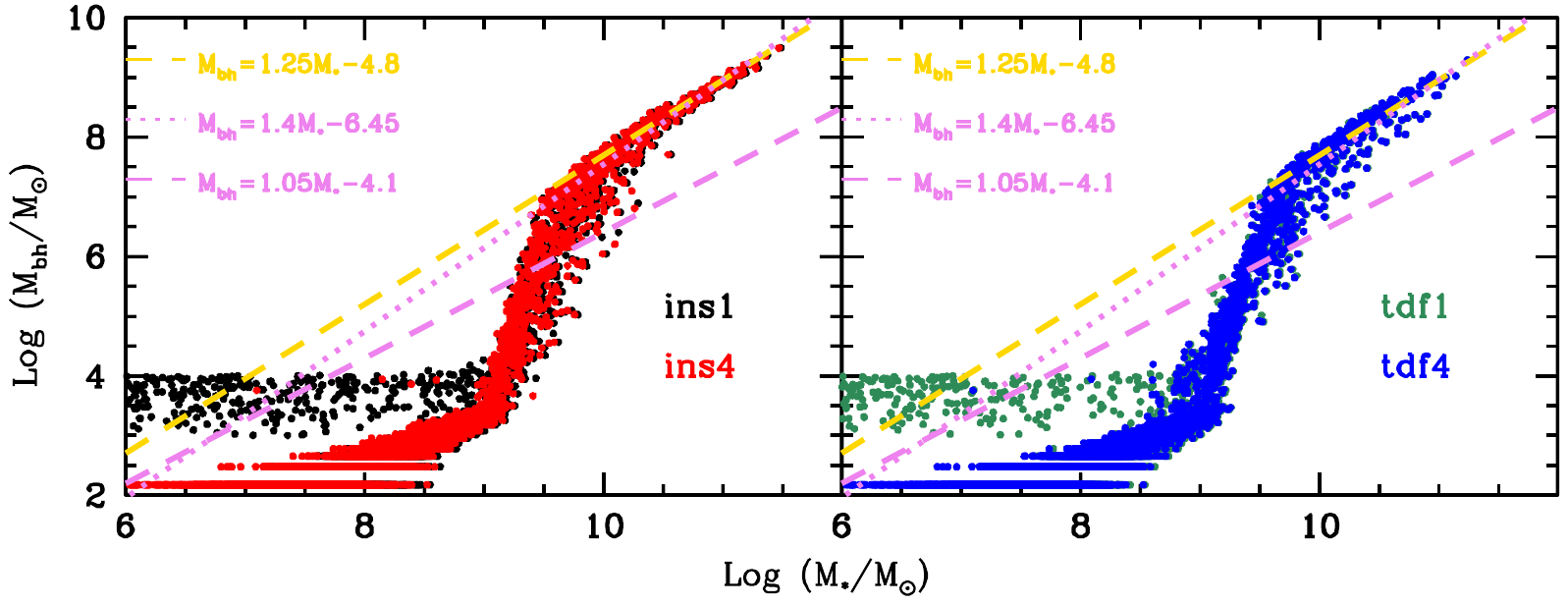}}
\caption{The black hole mass-stellar mass relation for $z\simeq 5$ for two models that bracket the expected range: Instantaneous mergers with/without UV feedback ({\it ins1} using black points and {\it ins4} using red points; left panel) and mergers after a merging timescale with/without UV feedback ({\it tdf1} using green points and {\it tdf4} using blue points; right panel). In both panels we show two relations derived using galaxies in the nearby Universe: $M_{bh}=1.4 M_*-6.45$ derived for high stellar mass ellipticals and bulges and $M_{bh}=1.05 M_*-4.1$ for moderate luminosity AGN in low-mass halos \citep{volonteri2016}, as marked. In each panel we also show the best-fit relation from our model for high stellar mass galaxies: $Log M_{bh} = 1.25 M_* - 4.8$. As seen, our theoretical model yields a non-linear scaling such that black holes in low-mass galaxies are ``stuck" at their initial mass; the BH masses of high-mass hosts, on the other hand, are strongly correlated with the stellar mass and are in excellent agreement with the results derived for lower-$z$ high stellar mass galaxies.   }
\label{nion_fnms} 
\end{figure*}

We also show the BH occupation fraction in Fig. \ref{fig_occu}. As shown, galaxies with a halo mass $M_h \gsim 10^{10.2}$ have an occupation fraction of 1 by $z \sim 6$. As expected, most of these are stellar black holes except DCBHs that dominate for the most massive halos. The black hole occupation fraction also shifts to progressively lower masses with increasing redshift. This is because of two reasons: first in our model, only starting leaves above $z =13$ are seeded with black holes; the increasing number of starting leaves forming at lower redshifts are devoid of any black holes. Secondly, low mass halos continually increase in mass with decreasing redshift. We note that our results are qualitatively in good agreement with those obtained from previous works \citep[e.g.][]{tanaka2009}. Finally we stress that the enhancement of the LW seen by any halos only depends on its bias at that redshift and we have ignored the impact of clustered sources that could enhance the LW intensity seen by halos in over-dense environments. Our results must therefore be treated as a lower limit on the DCBH number density and, hence, the Type 2+3 occupation fraction.

\subsection{The black hole-stellar mass relation}
Constraints on the relation between BHs and galaxies at high redshift are scant. In general, since the only confirmed BHs at these redshifts are those powering powerful quasars, the stellar mass of the host cannot be measured (not to mention the stellar velocity dispersion or bulge mass) because the light from the quasar over-shines the host galaxy.  The best estimates of the host properties for these powerful quasars are obtained through measures of the cold (molecular) gas properties in sub-mm observations, where a dynamical mass, based on the velocity dispersion of the gas and the radius  of the emitting region can be measured \citep[e.g.,][and references therein]{venemans2016,shao2017,decarli2018}. For these quasars, the BH to dynamical mass is skewed to values much larger than the ratio of BH to stellar or bulge mass in the local Universe. As discussed in \cite{volonteri2011} there are reasons to believe that such high mass ratios should not characterize the whole BH population. Beyond the Malmquist bias 
causing a more frequent selection of over-massive BHs in low-mass hosts  \citep{lauer2007,salviander2007}, only under-massive and low-accretion BHs can explain the lack of widespread AGN detections in LBGs. That BHs in low-mass galaxies are indeed expected to grow slowly and lag behind the host has now been confirmed in many numerical investigations \citep{dubois2015,habouzit2017,bower2017,angles2017}. Our implementation of BH growth includes a stunted growth in low-mass galaxies and we obtain a black hole-stellar mass relation in agreement with numerical investigations, a non-linear scaling where black holes in low-mass galaxies are ``stuck" at their initial mass \citep{habouzit2017,bower2017}. BHs in high-mass hosts, on the other hand, can be above the $z=0$ scaling, as shown in Fig.~\ref{nion_fnms}.

Quantitatively, we find that the BH mass-stellar mass relation is strongly correlated for high stellar mass ($M_* \gsim 19^{9.5}\msun$) galaxies and is 
best expressed by the relation $M_{bh}=1.25M_*-4.8$ at $z \simeq 5$; the relation flattens below such masses. Including the impact of the UVB ({\it ins4}) has no impact on this relation at the bright end. However, the suppression of gas mass in low-mass halos naturally results in lower black hole masses by as much as two orders of magnitude for a given stellar mass. As noted above in Sec. \ref{bhmf}, the inclusion of a delay in galaxy merging timescales results in a decrease in the mass of the most massive black holes (by about 0.8 dex) as seen from the right-hand panel of the same figure although it has no impact on the high-mass slope. Further, the results from {\it ins4} and {\it tdf4} are quite similar as also expected from the discussion in Sec. \ref{bhmf} above, yielding the lower-limit to the $M_{bh}-M_*$ relation. Finally, the best-fit relation derived for high stellar mass galaxies from our model is in excellent agreement 
with the relation $M_{\rm bh}=1.4 M_*-6.45$ derived for high stellar mass ellipticals and bulges in the nearby Universe \citep{volonteri2016}.

\section{{\it LISA} and GWs from the high-z Universe}
\label{lisa}
Now that we have shown the theoretical galaxy and BH properties to be in excellent agreement with observations, we can extend our calculations to the GWs expected from the mergers of such high-$z$ black holes. In this work, any merger falls into one of the following 3 categories: (i) {\it type1 - stellar black hole mergers}: mergers of two stellar BH seeds; (ii) {\it type 2 - mixed mergers}: mergers of a stellar BH seed with a DCBH, and (iii) {\it type 3 - DCBH-DCBH mergers}: extremely rare, these are mergers of two DCBH seeds. In the last category, we also include mergers of a DCBH with a mixed merger in the past. 

\begin{figure}
    \includegraphics[scale=0.5]{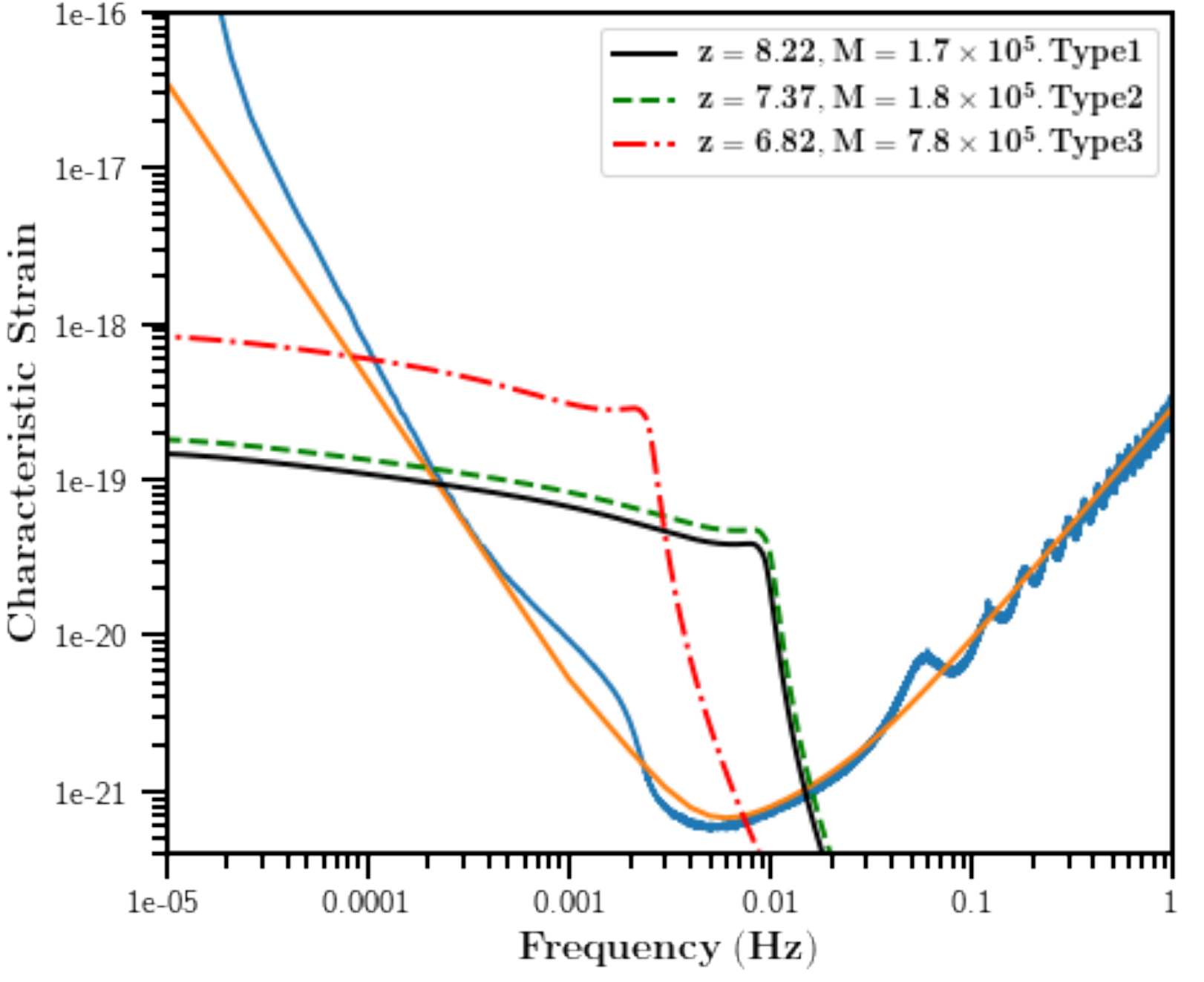}
   \caption{LISA sky-averaged dimensionless sensitivity curve ($\sqrt{f \times S_{\rm n}}$) as a function of frequency for 4 years of observation time: numerical calculation \citep[][blue line]{whitepaper} and analytical approximation  \citep[yellow line; model A2N2 from][]{klein2016}. The numerical curve accounts for the Galactic binary stochastic foreground noise that causes the ``bump" around $10^{-3}$ Hz. We also overplot the dimensionless characteristic strain $h_{\rm c} =  2 f \tilde{h}(f)$ for the averaged (in total mass and redshift) detected BH binaries in the 3 different type of mergers (using the ``light" DCBH seeds model) considered in this work. Note that the average detected binary is at an increasing redshift and has a decreasing total mass going from type $3$ to type $1$ mergers. }
    \label{fig:Sensitivity}
\end{figure}

A system with two black holes revolving around each other forms an accelerated mass quadrupole that causes emission of GWs at the expenses of orbital energy with a catastrophic outcome: as the binary emits GWs its semi-major axis shrinks (``inspiral" phase) until the two black holes merge and, after shedding any extra residual energy (``ringdown" phase), a newly born stationary BH forms. The GW signal increases in amplitude and frequency at an accelerated pace with the emission peaking at merger, i.e. roughly at the innermost stable circular orbit (ISCO). The peak frequency at the ISCO for a non-spinning black hole can be expressed as twice
\begin{equation}
    f_{\rm ISCO}=\frac{1}{6\sqrt{6}(2\pi)}\frac{c^3}{GM(1+z)}\,,
    \label{FISCO}
\end{equation}
where M is the total binary mass. Beyond the peak the signal is exponentially damped. Massive black holes ($M_{bh} > 10^{3} \msun$) at high redshifts emit at frequencies ($\ll 1$ Hz) much lower than the range of ground based GW detectors. To detect massive BHs through GWs much longer interferometric arms, of a million kilometers are needed, which can be only realised in space. In this section, we forecast the detection performance of the space-based European Space Agency (ESA) mission {\it LISA} for black hole binaries in the early Universe ($z >4$), in the evolutionary framework presented above. {\it LISA} is a space-based GW laser-interferometer, proposed to be launched in 2034, that consists of three spacecrafts in an equilateral triangle constellation. The interferometer's arms are proposed to be $2.5\times 10^6$ km in length resulting in an optimal frequency range between $\approx 1$ mHz to $0.1$Hz (Fig. \ref{fig:Sensitivity}). 

For each black hole merger, the optimized value of the signal to noise ratio (SNR) associated to the wave model is calculated based on the matched-filtering technique. By assuming the noise to be stationary and Gaussian with zero mean, the SNR is given by
\begin{equation}
    \left(\frac{S}{N}\right)^{2}=  4 \int_{f_{\rm min}}^{f_{\rm max}}\frac{|\Tilde{h}(f)|^{2}}{S_{\rm n}(f)}df\,,
    \label{eq:snr}
\end{equation}
where $|\Tilde{h}(f)|$ is the amplitude of the GW signal in frequency domain and $S_{\rm n}(f)$ is the noise power spectra density (PSD) function. Here $f_{\rm min}$ is the binary frequency when it is first observed (i.e. at $t=0$) and $f_{\rm max}$ is either its frequency at the end of the mission's lifetime or at merger, which ever happens first. In the following sections, we detail the calculation of  the signal ($h(f)$; Sec. \ref{lisa_signal}) and the noise ($S_{\rm n}$) and integration limits (Sec. \ref{lisa_noise}). We end by presenting our results in Sec. \ref{lisa_results}. 

\subsection{The GW signal }
\label{lisa_signal}

GW detectors generally work with time-dependant scalars, $h(t)$, as their output. The scalar describes the changes in the detector after the passage of waves. In case of laser interferometers $h(t)$ represents the phase shift of the laser beam (or equivalently, the change in the detector's arm length) and can be expressed as
\begin{equation}
    h(t)=F_{\times}h_{\times}(t)+F_{+}h_{+}(t)\,,
\end{equation}
where $h_{\times}(t)$ and $h_{+}(t)$ are the GW polarizations. Further, $F_{\times}$ and $F_{+}$ are detector's pattern functions that depend both on the properties of the detector as well as the position of the source in the sky. After averaging the signal over the sky position and transforming it into the frequency domain we obtain
\begin{equation}
    |\Tilde{h}(f)|^{2}=|\Tilde{\mathcal{A}}(f)|^{2}\times|Q^{2}|\,,
\end{equation}
where  $\Tilde{\mathcal{A}}(f)$ is the wave amplitude in the frequency domain and $Q$ is a geometrical factor containing information about the pattern functions. For our choice of the detector's configuration we use $Q=\frac{2}{10}\sqrt{3}$, which accounts for averaging over sky position and binary inclination \cite[see][]{whitepaper}.

The calculation of $\Tilde{\mathcal{A}}(f)$ should in principle be performed with a fully relativistic (NR) numerical code. However, such calculations are numerically expensive and, in fact, only necessary for modelling the highly relativistic end of the inspiral phase and merger. The inspiral phase, where orbital velocities are much lower than the speed of light, can instead be satisfactorily reproduced with an analytical post-Newtonian (PN) formalism. These considerations inspired the so-called ``phenomenological models" that give a complete analytical wave model by matching the PN and NR waveforms in the region where the PN approximation breaks down. To calculate $\Tilde{\mathcal{A}}(f)$, we model the waveform with the phenomenological model ``PhenomC", which has the advantage of producing the waveform directly in frequency domain, convenient for data-analysis applications \citep[for a detailed description of the code see][]{phenomC}.

\begin{figure*}
\center{\includegraphics[scale=0.65]{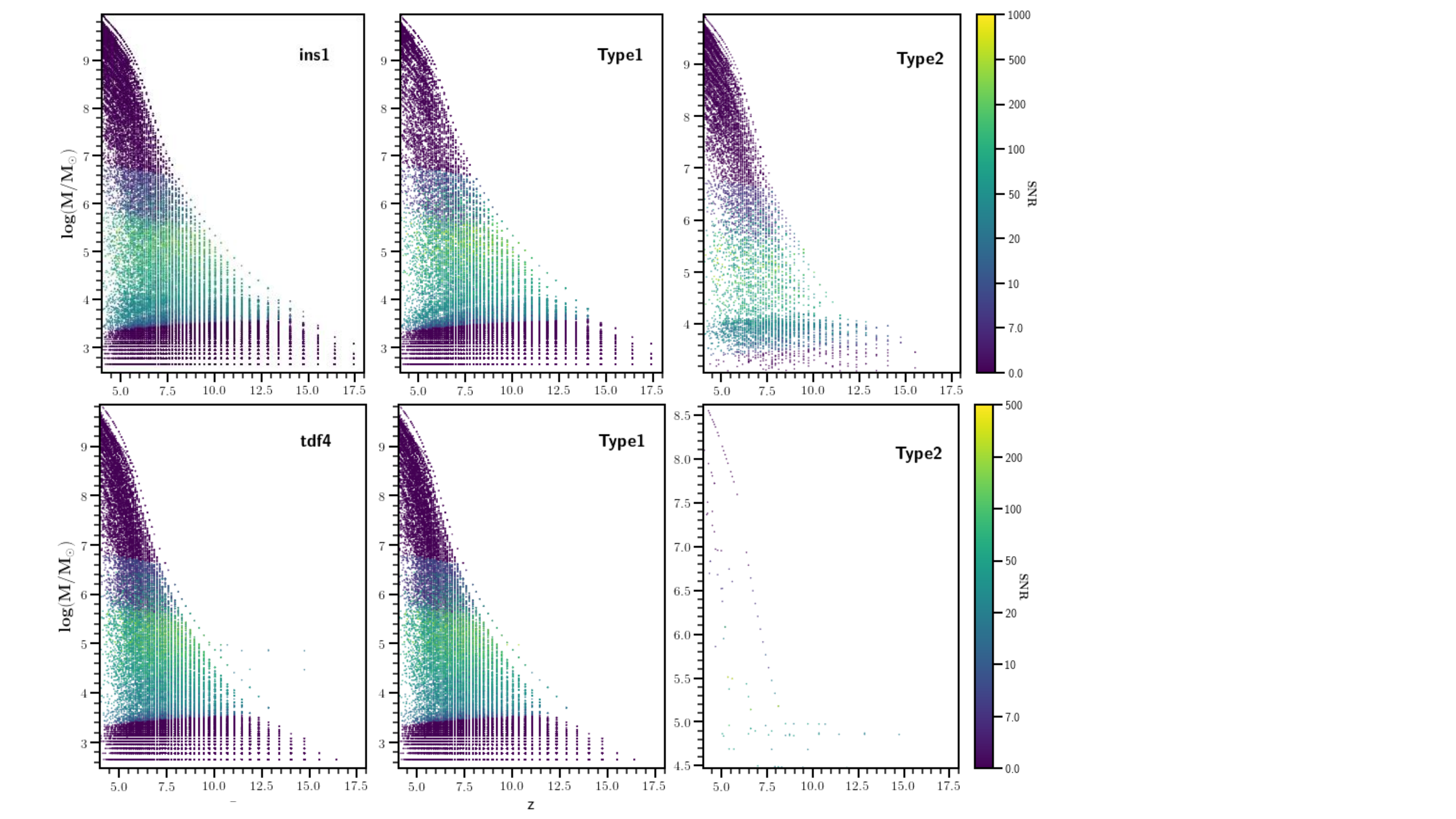}}
\caption{The Signal to Noise Ratio (SNR) as a function of intrinsic total binary mass and $z$ for 4 year mission duration. The columns from left to right show results for all mergers, type 1 mergers (mergers of two stellar BHs) and type 2 mergers (mergers of a stellar BH and a DCBH); there are no detections of type 3 black holes (mergers of two DCBHs). As marked, the upper and lower rows correspond to results for models {\it ins1} and {\it tdf4} with ``light" DCBH seeds, respectively. The {\it LISA} detectability window is such that binaries with SNR $> 7$ have redshifts between $z =5-13$ and a total mass between $M \simeq 10^{3.5-5.6}\msun$ with the exact value depending on the model and merger type. The characteristic strain for the average binary is traced in Fig. \ref{fig:Sensitivity}.}
\label{trend} 
\end{figure*}
\subsection{The instrumental and source noise}
\label{lisa_noise} 
We numerically calculate the sky-averaged noise PSD, $S_{\rm n}$, for {\it LISA} using the {\it LISA}-consortium simulator, that takes into account different instrumental noises as well as the stochastic background from unresolved Galactic binaries. The most notable contribution to the latter comes from Galactic white dwarf binaries that {\it LISA} is unable to resolve individually \citep[e.g.][]{DWD}. The number of these sources is expected to decrease as the mission progresses and a larger number of foreground sources are detected and removed. {\it LISA}'s sky-averaged sensitivity curve ($\sqrt{f \times S_{\rm n}}$) adopted in this paper for the SNR calculation corresponds to a 4-year observing time and is presented (using the blue line) in Fig. \ref{fig:Sensitivity}.
For convenience, the frequency limits of the integral (Eqn.~\ref{eq:snr}) are instead calculated adopting an analytical fit to {\it LISA}'s PSD of the form
\begin{equation}\label{PSD}
\begin{split}
    S_{n}(f)&=\frac{20}{3}\frac{4S_{\rm n,acc}(f)+S_{\rm n,sn}(f)+S_{\rm n,omn}(f)}{L^{2}}\\
    &\times \left[1+\left(\frac{f}{\frac{0.41c}{2L}}\right)^{2} \right]\,,
    \end{split}
\end{equation}
(yellow solid line in Fig. \ref{fig:Sensitivity}) from \cite{klein2016}. In the above equation, $L$ is the detector arm length. Further, $S_{\rm n,acc}$, $S_{\rm n,sn}$ and $S_{\rm n,omn}$ are the noise components due to low-frequency acceleration, shot noise and other measurement noise, respectively. Instead of performing a formal fit to the numerical curve in order to estimate the noise parameters, we adopt the following values from those reported in \cite{klein2016}:
\begin{equation}
    \begin{split}
        &S_{\rm n,acc} =\frac{9\times10^{-30}}{(2\pi f)^{4}}(1+\frac{10^{-4}}{f})\,\,[\rm{m^{2}Hz^{-1}}],\quad \\
        &S_{\rm n,sn} =2.22\times 10^{-23}\,\,[\rm{m^{2}Hz^{-1}}],\quad \\
        &S_{\rm n,omn} =2.65\times 10^{-23}\,\,[\rm{m^{2}Hz^{-1}}],
    \end{split}
\end{equation}
corresponding to a $L= 2$ Mkm arm length \citep[model A2N2L6;][]{klein2016}. Both the numerical and analytical curve in Fig. \ref{fig:Sensitivity} are calculated for the current LISA design, that presents three spacecrafts connected by 6 links.

A visual comparison between our analytical (Eqn. \ref{PSD}) and numerical curves shows a sufficiently close match for our purposes for frequencies $> 10^{-4} ~{\rm Hz}$. We note that the analytical curve does not account for a stochastic background (the bump around $10^{-3} ~{\rm Hz}$). This omission allows low-mass black hole binaries ($\sim 10^3 M_{\odot}$), that cross the sensitivity curve near its bottom, to stay longer in the observed band. However, we will show that, even with this extra integration time, their SNR is never above the detection threshold. Finally, we over-plot the signals from representative detected merger events. The examples considered in this figure are the average (in total mass and redshift) detected binary for each of the 3 types of mergers considered in this work. Their tracks cross the {\it LISA} sensitivity curve around $f_{\rm min} \approx {\rm ~a ~few} ~10^{-4} ~{\rm Hz}$ where the analytical and numerical sensitivity curves match extremely well.

\begin{table*}
\centering
\caption {Total number of {\it LISA} detections expected for a SNR $ >7$ at $z>4$ over a 4-year duration of the mission for the two models the bracket the upper and lower limits of the physical parameter space: {\it ins1} and {\it tdf4} for both light and heavy DCBH seeds. We show results for the three different types of BH mergers explained in Sec. \ref{lisa_noise}: (i) {\it type1 - stellar black hole mergers}: mergers of two stellar black hole seeds; (ii) {\it type 2 - mixed mergers}: mergers of a stellar black hole seed with a DCBH, and (iii) {\it type 3 - DCBH-DCBH mergers}: extremely rare, these are mergers of two DCBH seeds. While columns 2-5 show results for all, type 1, 2 and 3 mergers, respectively, for a SNR$>7$, columns 6-9 show results for the same quantities without imposing any SNR cut.}
\centerlast
\begin{tabular}{ |c|c|c|c|c|c|c|c|c|}
 \hline
  Model &  All$_7$ &  Type 1$_7$ & Type 2$_7$ & Type 3$_7$ & All$_0$ &  Type 1$_0$ & Type 2$_0$ & Type 3$_0$ \\
   \hline
     ins1 & 19.8 &   13 & 6.8 & 0.05  & 300.3 & 288.3 & 11.8 & 0.15 \\
  tdf4 &  12.5 & 12.1 & 0.4 & 0  & 247.7  & 247.0  & 0.62 & 0.01 \\
  ins1 (heavy) &  23.3 & 13 & 10.3 &   0.04  & 300.3 & 288.3 & 11.8 & 0.15 \\
  tdf4 (heavy) & 12.5 & 12.1 & 0.4 & 0  & 247.7 & 247.0 & 0.62 & 0.01 \\
  \hline
 \end{tabular}
  \label{table_GW}
\end{table*} 

\subsection{{\it LISA} detectability of GW from the high-z Universe}
\label{lisa_results}
To confidently claim detection, the SNR of an event must be above a critical value. Here we adopt the typical {\it LISA} threshold of ${\rm SNR=7}$. Each row in Fig. \ref{trend} represents the calculated SNR values for all simulated binaries in a given model as a function of their total intrinsic mass and the redshift. Starting with the fiducial model ({\it ins1}; top panels), BH mergers become detectable once they reach masses of $\sim 10^{4}$ $M_{\odot}$ at $z \lsim 13$. As masses grow with time, these systems can reach SNR values as high as $\sim 1000$ for a total BH mass around $10^{5} M_{\odot}$ below $z \sim 11$. Binaries with SNR $> 7$ appear in the redshift range $z \simeq 5-13$ and range in total mass between $M \simeq 10^{3.5-5.6}\msun$. Allowing a precise estimation of parameters such as distance, sky localisation and chirp mass, these mass and $z$ ranges will therefore be best probed using GWs. Finally, as the black holes grow above $\sim 10^{6} \msun$, the SNR decreases as the emitted GW signal shifts to lower frequencies, and above $\sim 10^{7} \msun$ it goes out of the detectability window. While the results remain quite similar for the {\it tdf4} model, a delay in the merger timescales results in a severe reduction in the number of type 2 mergers as shown from the lower right-most panel of the same figure. Moreover, the detectability window around $10^{4-6} \msun$ shifts to slightly lower redshifts. We end by noting that type 2 mergers are rarer than type 1 in both models, with type 3 mergers being the rarest as expected (see Table \ref{table_GW}) - this is why type 3 mergers are not plotted here. 

\begin{figure*}
\center{\includegraphics[scale=0.99]{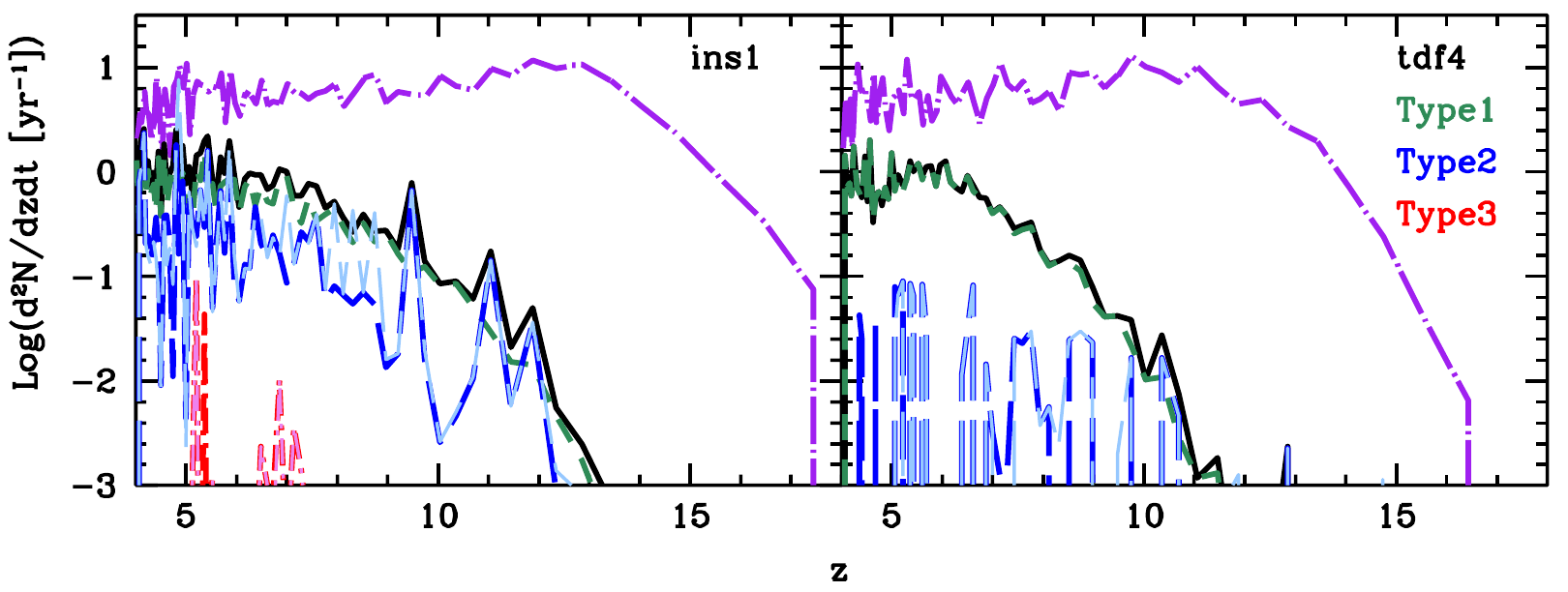}}
\caption{The BH merger event rate (per year) expected as a function of redshift for two models that bracket the physical range probed: {\it left panel}: ins1 and {\it right panel}: tdf4. In each panel, the dot-dashed purple line shows the results for all mergers (without any cut in signal to noise ratio) while the solid black line shows the results for all mergers using a value of SNR$>7$. The latter is deconstructed into the contribution from (SNR$>7$) type1 (green dashed line), type2 ``light DCBH" seed (dark blue dashed line) and type3 ``light DCBH" seed (red dashed line) mergers. Further, the long-dashed light blue line and dot-dashed pink line show results for mergers with SNR$>7$ using a heavier DCBH seed mass of $10^{4-5}\msun$ for type 2 and type 3 mergers, respectively. These results are in general agreement with those used for {\it LISA} calculations \citep[e.g. Fig. 3][]{klein2016}.}
\label{eventz}
\end{figure*}

We now discuss the yearly high-$z$ event detection rate expected from {\it LISA} using a SNR$>7$. Once the BH merger rate density (per unit comoving volume) of events with SNR$ >7$ at a given z, $N_{\rm com}(z)$, is obtained, we convert this into the expected number of mergers per year $d^2N/dzdt$ as \citep{haehnelt1994, arun2009}
\begin{equation}
\frac{d^2N}{dz dt} =  4 \pi c N_{com}(z) \bigg(\frac{d_L(z)}{(1+z)}\bigg)^2 \, [yr^{-1}],
\end{equation}
where $d_L(z)$ is the luminosity distance at $z$. The results of this calculation are shown in Fig. \ref{eventz}. As shown, the fraction of {\it LISA detectable} events rises with decreasing redshift from about 1/400 at $z \simeq 13$ to about 1/10 by $z \simeq 8$ to as high as 1/4 by $z \simeq 5$. As expected, most of these events are type 1 mergers. Quantitatively, by $z \simeq 4$, roughly 66\% of detectable mergers are type 1 with about 32\% being type 2 mergers with type 3 mergers only contributing 0.3\% to the total number. While the qualitative behaviour is quite similar in the {\it tdf4} case, given the slower BH mass growth, type 1 mergers significantly increase (contributing about 96\% to the cumulative event rate by $z \simeq 4$) while the contribution of type 2 mergers falls to roughly 3\%. Crucially, we do not find any type 3 mergers above the detection limit in this case.

We find that considering the ``heavy DCBH seed" model leads to a slight change in these numbers for the {\it ins1} case: while the cumulative contribution of type 1 mergers drops slightly to $52\%$, this is compensated by an increase (to 47\%) in the cumulative number of detectable type 2 mergers while the number of type 3 mergers remain unchanged. This heavier seed model, however, has no impact on the results from the {\it tdf4} model 

The {\it total} number of detections per model and merger type for the {\it LISA} mission (over 4 years) are summarised in Table \ref{table_GW}. The model {\it ins1} with ``heavy DCBH seeds" yields the highest total detection number of $\sim 23$ events comprising of $\sim 13$ type 1 and $\sim 10$ type 2 mergers. These numbers reduce slightly to about 20 total events comprising of 13 type 1 and 7 type 2 mergers using the ``light DCBH seed" model. In contrast, only a dozen events (all of type 1) are expected using model {\it tdf4}; as expected from the discussion above, the DCBH seed mass has no bearing on these results.

\begin{figure*}
\center{\includegraphics[scale=0.99]{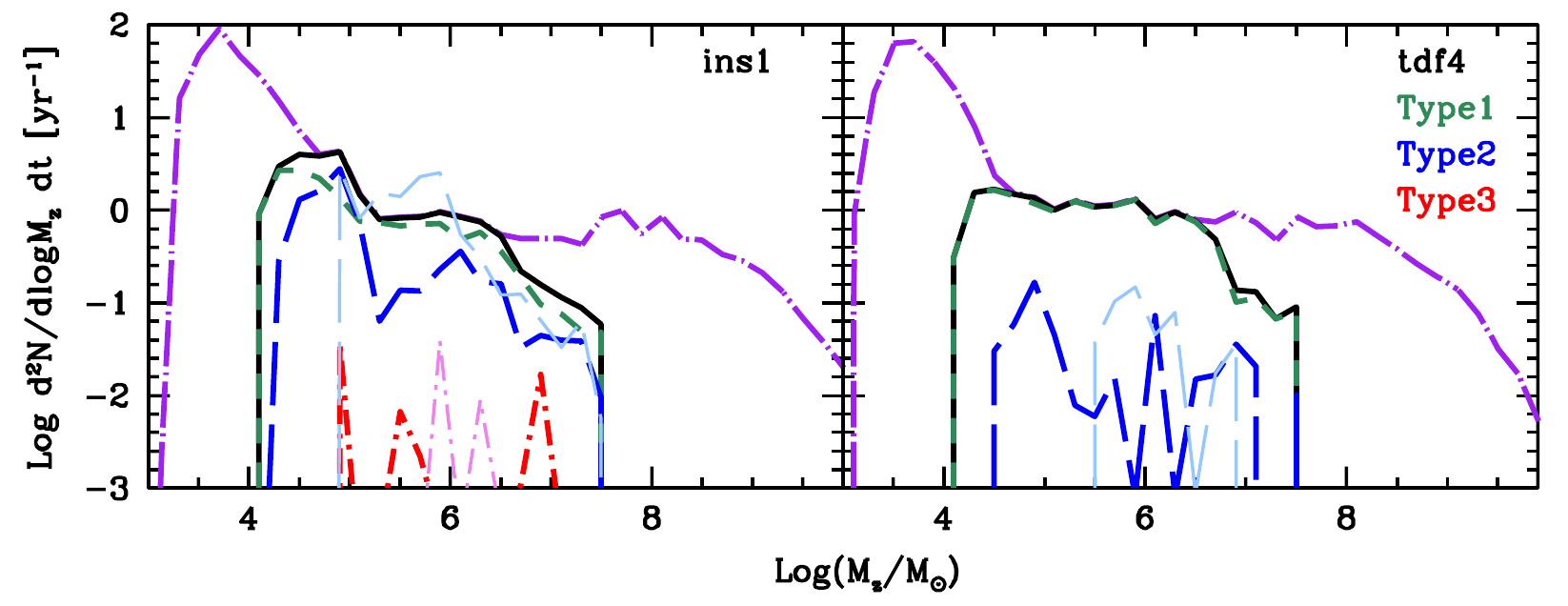}}
\caption{The BH merger event rate (per year) as a function of the redshifted BH mass ($M_z = M_{\rm bh}(1+z)$). The lines show the same models as noted in Fig \ref{eventz}.}
\label{eventm}
\end{figure*}

We also calculate the event rate in terms of the redshifted merged mass, $M_z = M(1+z)$, such that 
\begin{equation}
\frac{d^2N}{dM_z dt} =  4 \pi c N_{com}(M_z) \bigg(\frac{d_L(z)}{(1+z)}\bigg)^2 \, [yr^{-1}]. 
\end{equation}
The results of this calculation, presented in Fig. \ref{eventm}, clearly show the {\it LISA} detectability preference for BH masses ranging between $10^4-10^7 \msun$ for type 1 and type 2 mergers for both the {\it ins1} and {\it tdf4} models. Type 3 mergers, instead, are detectable in the mass range $10^{5-7}\msun$ in the ``light" DCBH seed model while being undetectable in the {\it tdf4} model. Moving on to the ``heavy DCBH seed model", while the mass range remains unchanged for type 1 mergers, the range for both type 2 and type 3 mergers decreases: while the former range between $10^{5-7}\msun$ for both the {\it ins1} and {\it tdf4} models, the type 3 range lies in the very narrow range of $10^{5.5-6.5}\msun$ for the {\it ins1} case; as expected, the number of mergers of each type in each model are similar to the cumulative numbers quoted above. Practically, however, it would be difficult to distinguish between these different seeding models purely from the detected mass function given all types of merger reside in the same mass range between $10^{4-7}\msun$. 

Finally, we provide a comparison of our expected event rates with those available in the literature: starting with heavy seeds, all previous studies used DCBH models based on ``dynamical" instabilities, of the type advocated by \citet{begelman2006}, \citet{lodato2006} and \citet{2010MNRAS.409.1022V}. In this study we have focused on the currently favored (at least by the first star community) ``thermodynamical" models, that require a high level of LW background for the formation of seeds. As shown in this paper and in \citet{habouzit2016} this model results in much rarer seeds. We find that \citet{klein2016} predict 3.9 mergers/year at $z>4$ in the Q3-d model based on \citet{lodato2006}; note that using the same seeding model as \citet{klein2016}, \citet{bonetti2018}  obtain results consistent with previous literature. Further, \citet{sesana2007} predict 2.2 mergers/year at $z>4$  in the model based on \citet{begelman2006}, while the LW-based model explored in this paper yields 0.0025-0.035 mergers/year at $z>4$ as shown in Table \ref{table_GW}. A comparison with \citet{ricarte2018}, who also use a model based on \citet{lodato2006} and do not include a LW condition, is more difficult because they show only events with SNR$>5$. Using this SNR cut, we find that the peak in the rates for heavy seeds is similarly broad and covers a similar redshift range when comparing our results to theirs although they predict a larger number of events: their peak rate is between $0.5-5$ events/year while our peak rate goes from $ \sim 0.05 -0.25$ events/year. To summarise, our lower merger rates for heavy seeds, compared to previous works, is what should be expected for a model that predicts extremely rare seeds. This is the effect of adding the condition on the LW background that previous models had not included. 

For light (popIII) seeds \citet{klein2016} predict 146.3 mergers/year at $z>4$ (although they extrapolate to 2x this rate in their Table 1 and related text) and \citet{sesana2007} predict 57.7 mergers/year at $z>4$. Our model predicts between 62.0 and 75.1 mergers/year at $z>4$ as shown in Table \ref{table_GW}. When comparing to \citet{ricarte2018}, again the peak in the rates for light seeds is similarly broad and covers a similar range in redshift but the value of the peak rates are lower in our case. In particular, our type 1 peak rate is ~0.75 event/yr in the optimistic ({\it ins1}) and pessimistic ({\it tdf4}) models, while in Fig. 9 their peak rates lie between $\sim 5-20$ events/year. While our merger rate for light seeds is well within the expectations of the literature, as we made similar assumptions, the results being on the lower side are likely because of the resolution of our merger trees: for instance, \citet{ricarte2018} have a mass resolution of $\sim 10^6 \msun$, while \citet{klein2016} follows \citet{barausse2012} who follows \citet{volonteri2003} \citep[whose trees are used for][] {sesana2007} in having a resolution dependent on the halo mass at z=0, reaching $10^5 \msun$ for halos with mass $<4.10^{12} \msun$ at z=0 and up to $10^7 \msun$ for halos with mass $10^{15}\msun$ at z=0.

\section{Conclusions and discussion}
\label{conclusions}

In this work, we have included the impact of BH seeding, growth and feedback, into our semi-analytic model, {\it Delphi}. Our model now jointly tracks the build-up of the dark matter halo, gas, stellar and BH masses of high-$z$ ($z \gsim 5$) galaxies. We remind the reader that our star formation efficiency is the minimum between the star formation rate that equals the halo binding energy and a saturation efficiency. In the same flavour, the BH accretion at any time-step is the minimum between the BH accreting a certain fraction of the gas mass left-over after star formation, up to a fraction of the Eddington limit: while high-mass halos can accrete at the Eddington limit, low-mass halos follow a lower efficiency track. We explore a number of physical scenarios using this model that include: {\it (i)} two types of BH seeds (stellar and those from Direct Collapse BH; DCBH); {\it (ii)} the impact of reionization impact; and {\it (iii)} the impact of instantaneous versus delayed galaxy mergers on the baryonic growth.

We show that, using a minimal set of mass and $z$-independent free parameters, our model reproduces all available data-sets for high-$z$ galaxies and BH including the evolving (galaxy and AGN) UV LF, the SMD and the BHMF. Crucially, our model naturally yields a BH mass-stellar mass relation that is tightly coupled for high stellar mass ($M_* \gsim 10^{9.5}\msun$) halos; lower-mass halos, on the other hand, show a stunted BH growth. Interestingly, while both reionization feedback and delayed mergers have no impact on the UV LF, the SMD is more affected by reionization feedback as compared to delayed mergers. 

We then use this model, bench-marked against all available high-$z$ data, to predict the merger event rate expected for the {\it LISA} mission. We find that {\it LISA}-detectable binaries (with SNR $> 7$) appear in the redshift range $z \simeq 5-13$ and range in total mass between $M \simeq 10^{3.5-5}\msun$. While type 1 mergers (of two stellar BHs) dominate in all the scenarios studied, type 2 mergers (merger of a stellar BH and a DCBH) can contribute as much as 32\% to the cumulative event rate by $z \sim 4$ in the fiducial ({\it ins1}) model. However including the impact of reionization feedback and delayed mergers ({\it tdf4} model) results in a lower BH growth with type 2 mergers contributing only 3\% to the cumulative event rates. Using heavier DCBH seeds results in a larger number of type 2 mergers becoming detectable with {\it LISA} whilst leaving the results effectively unchanged for the {\it tdf4} model.  

Quantitatively, the model {\it ins1} with ``heavy DCBH seeds" yields the highest total detection number of $\sim 23$ events comprising of $\sim 13$ type 1 and $\sim 10$ type 2 mergers. These numbers reduce slightly to about 20 total events comprising of 13 type 1 and 7 type 2 mergers using the ``light DCBH seed" model. In contrast, only a dozen events (all of type 1) are expected using model {\it tdf4} and the DCBH seed mass has no bearing on these results. 

We end with a few caveats. Firstly, given that we do not consider (the realistic case of) recoil and BH ejection form the host halos, all BHs remain bound to halos. Secondly, the enhancement of the LW seen by any halos only depends on its bias  at that redshift. This effectively means that we ignore the impact of the local environment on the LW intensity seen by any halo, and this may lead to an increase in seed formation and mergers in more biased regions.  Thirdly, we have not included BH seeds from stellar dynamical channels which have a milder metallicity dependence and should have a number density intermediate between SBHs and DCBHs \citep[e.g.,][]{2012MNRAS.421.1465D,2014MNRAS.442.3616L}; DCBH models that are metallicity-independent can also provide an additional channel increasing the BH merger rate over cosmic time \citep{2010MNRAS.409.1022V,2014MNRAS.437.1576B}. Finally, we have used a very crude mode for reionization feedback that ignores the patchiness of reionization - in our model, halos either remain unaffected by the UVB or halos below a certain chosen virial velocity have all of their gas mass completely photo-evaporated. We aim to address each of these intricacies in detail in future works.

\section*{Acknowledgments} 
PD acknowledges support from the European Research Council's starting grant ERC StG-717001 (``DELPHI"). PD and OP acknowledge support from the European Commission's and University of Groningen's CO-FUND Rosalind Franklin program. MV acknowledges funding from the European Research Council under the European Community's Seventh Framework Programme (FP7/2007-2013 Grant Agreement no.\ 614199, project ``BLACK''). Finally, PD thanks A. Mazumdar for his scientific inputs which have greatly added to the paper.


\bibliographystyle{mn2e}
\bibliography{gw_rv}

\label{lastpage} 
\end{document}